\documentclass{article}

\usepackage{amssymb, amsfonts}

\usepackage{amsthm}
\newtheorem{theorem}{Theorem}
\newtheorem{lemma}{Lemma}
\newtheorem{proposition}{Proposition}
\newtheorem{corollary}{Corollary}

\theoremstyle{definition}

\newtheorem{remark}{Remark}

\newcommand{\qedi}{ }

    \newcommand {\R} {\ensuremath{\mathbb{R}}}
    
    \newcommand {\N} {\ensuremath{\mathbb{N}}}

    \newcommand {\mr} {\mathrm}

    \newcommand {\C} {\mathbb C}

  \newcommand{\commut}[2]{\left[ #1 , #2 \right]}
  \newcommand{\antcom}[2]{\left[ #1 , #2 \right]_+}
  \newcommand{\eps}{\varepsilon}
 \newcommand{\Or}{\mathcal{O}}
  \newcommand{\e}{\mathrm{e}}
  
  \newcommand{\tRe}{\mathrm{Re}}
  \newcommand{\tIm}{\mathrm{Im}}
\newcommand{\id}{\mathrm{id}}
\newcommand{\D}{\mathrm{d}}

\newcommand{\I}{\mathrm{i}}
\newcommand{\op}{{n_\eps}}
\newcommand{\od}{{\rm}}

\newcommand{\alp}{\renewcommand{\labelenumi}{{\rm(\alph{enumi})}}}

\begin{document}

\title{\LARGE \bf Precise coupling terms in adiabatic\\ quantum evolution}

\author{Volker Betz\\ {\normalsize Institute for Biomathematics and Biometry, GSF
Forschungszentrum,}\\ {\normalsize Postfach 1129, D-85758
Oberschlei{\ss}heim,
Germany}\\{\normalsize volker.betz@gsf.de}\\[5mm] Stefan Teufel\\{\normalsize
Mathematics Institute, University of
Warwick,}\\ {\normalsize Coventry CV4 7AL, United
Kingdom}\\{\normalsize teufel@maths.warwick.ac.uk}}

\date{\today}

\maketitle

\begin{abstract}
It is known that for multi-level time-dependent quantum systems
one can construct superadiabatic representations in which the
coupling between separated  levels is exponentially small in the
adiabatic limit. For a family of two-state systems with
real-symmetric Hamiltonian we construct such a superadiabatic
representation and explicitly determine the asymptotic behavior of
the exponentially small coupling term. First order perturbation
theory in the superadiabatic representation then allows us to
describe the time-development of exponentially small adiabatic
transitions. The latter result rigorously confirms the predictions
of Sir Michael Berry for our family of Hamiltonians and slightly
generalizes a recent mathematical result of George Hagedorn and
Alain Joye.
\end{abstract}

\section{Introduction and main result}

The decoupling of slow and fast degrees of freedom in the
adiabatic limit is at the basis of many important approximations
in physics, as, e.g., the Born-Oppenheimer approximation in
molecular dynamics and the Peierls substitution in solid state
physics. We refer to \cite{BMKNZ,Te} for recent reviews.
Generically the decoupling is not exact and a coupling which is
exponentially small in the adiabatic parameter remains. However,
this small coupling has important physical consequences, as it
makes possible, e.g., non-radiative decay to the ground state in
molecules. Since Kato's proof from 1950 \cite{Ka} the adiabatic
limit of quantum mechanics was considered also as a mathematical
problem, with increased activity during the last 20 years. Some of
the landmarks are \cite{Nen1,ASY,JoPf1,Nen2,HaJo}.

We consider  a two-state time-dependent quantum system described
by the Schr\"o\-dinger equation
\begin{equation} \label{H}
\big( \I \eps \partial_t -H(t)\big)  \psi(t) = 0
 \end{equation}
 in the {\em adiabatic limit} $\eps\to 0$.
For the moment we take the Hamiltonian $H(t)$ to be the
real-symmetric $2 \times 2$-matrix
\[
H(t) = \rho(t)\left( \begin{array}{cc} \cos\theta(t) &\,\, \sin\theta(t) \\
                                  \sin\theta(t) & \,-\hspace{-1pt} \cos\theta(t) \end{array}
                                  \right)\,.
\]
The eigenvalues of $H(t)$ are $\pm\rho(t)$ and we assume that the
gap between them does not close, i.e.\ that $2\rho(t)\geq g>0$ for
all $t\in\R$.

As to be explained, even for this simple but prototypic problem
there are open mathematical questions. In order to explain the
concern of our work, namely  the time-development of the
exponentially small adiabatic transitions, let us briefly recall
some important facts about (\ref{H}).
 Let $U_0(t)$ be the orthogonal matrix that diagonalizes
 $H(t)$, i.e.
\begin{equation} \label{U_0}
 U_0(t) = \left( \begin{array}{cc} \cos(\theta(t)/2) & \sin(\theta(t)/2)
 \\ \sin(\theta(t)/2) & -\cos(\theta(t)/2)
\end{array} \right)\,.
\end{equation}
 Then the Schr\"odinger equation in the {\em adiabatic representation} becomes
\[
U_0(t)\,\big( \I \eps \partial_t -H(t)\big)\,U_0^{\ast}(t)\,
U_0(t) \psi(t) =: \big( \I\eps\partial_t - H^{\rm a}_\eps(t)\big)
\psi^{\rm a}(t) = 0
\]
with
\[
H^{\rm a}_\eps(t) = \left(\begin{array}{cc} \rho(t)&
\frac{\I\eps}{2}\theta'(t)\\[2mm]-\frac{\I\eps}{2}\theta'(t)&-\rho(t)\end{array}\right)\qquad
\mbox{and}\qquad \psi^{\rm a} (t) = U_0(t) \psi(t)\,.
\]
Here and henceforth, primes denote time derivatives. First order
perturbation theory in the adiabatic representation (cf.\ proof of
Corollary~\ref{PropCor}) and integration by parts yields the
adiabatic theorem \cite{BoFo,Ka}: The off-diagonal elements of the
unitary propagator $K^{\rm a}(t,s)$ in the adiabatic basis, i.e.\
the solution of
\[
\I\eps\partial_t K^{\rm a}_\eps(t,s) = H_\eps^{\rm a}(t)
K_\eps^{\rm a}(t,s) \,,\qquad K_\eps^{\rm a}(s,s) = {\rm id}\,,
\]
vanish
in the limit $\eps\to 0$. More precisely, let
\begin{equation}\label{Pdef}
P_+= \left(\begin{array}{cc}1&0\\0&0\end{array}\right)\,,\qquad
P_-= \left(\begin{array}{cc}0&0\\0&1\end{array}\right)\,,
\end{equation}
which project onto the adiabatic subspaces in the adiabatic
representation. Then
\begin{equation}\label{AdiabaticThm}
\|\, P_- \, K^{\rm a}_\eps(t,s)\, P_+ \| = \Or(\eps)\,.
\end{equation}
Therefore the transitions between the adiabatic subspaces are
$\Or(\eps)$. This bound is optimal in the sense that in regions
where $\theta(t)$ is not constant the leading order term in the
asymptotic expansion of $P_- \, K^{\rm a}_\eps(t,s)\, P_+$ in
powers of $\eps$ is proportional to $\eps$.

However, if $\lim_{t\to\pm\infty}\theta'(t)=0$ then in the
scattering limit  the transitions between the adiabatic subspaces
are much smaller: if the derivatives of $\theta\in C^\infty(\R)$
decay sufficiently fast, then for any $n\in\N$
\begin{equation}\label{Aeps}
\mathcal{A}(\eps):=\lim_{t\to\infty}\|\, P_- \, K^{\rm
a}_\eps(t,-t)\, P_+ \| = \Or(\eps^n)\,.
\end{equation}
If $\theta$ is analytic in a suitable neighborhood of the real
axis, then transition amplitudes are even exponentially small,
$\mathcal{A}(\eps) = \Or(\e^{-c/\eps})$ for some constant $c$
depending on the width of the strip of analyticity, see
\cite{JoPf1,Ma}.

It is well understood, see \cite{Le,Ga,Nen1}, how to reconcile the
apparent  contrariety between the smallness of the final
amplitudes in (\ref{Aeps}) and the optimality of
(\ref{AdiabaticThm}): the adiabatic basis is not the optimal basis
for monitoring the transition process. For any $n\in \N$ there
exist unitary transformations $U_\eps^n(t)$ such that the
Hamiltonian in this $n^{\rm th}$ superadiabatic representation
takes the form
\begin{equation}\label{nthorderSAB}
H^{n}_\eps(t) = \left(\begin{array}{cc} \rho^n_\eps(t)& c_\eps^n(t)\\
\overline{c}_\eps^n(t)&-\rho^n_\eps(t)\end{array}\right)\,\,\,\mbox{with}\,\,\,
\rho^n_\eps(t) = \rho(t) +\Or(\eps^2)\,\,\,\mbox{and}\,\,\,
|c_\eps^n(t)| = \Or(\eps^{n+1})\,.
\end{equation}
 In the $n^{\rm th}$ superadiabatic basis the off-diagonal components of
 the propagator and hence also the transitions are of order
$\Or(\eps^n)$, i.e.\ there are constants $C_n$ such that
\begin{equation}\label{SuperAdiabaticThm}
\|\, P_- \, K^{n}_\eps(t,s)\, P_+ \| \leq C_n \eps^{n}\,.
\end{equation}
In the scattering regime, where $\theta(t)$ becomes constant, the
superadiabatic bases agree with the adiabatic basis, i.e.\
$\lim_{t\to\pm\infty} U_\eps^n(t) = U_0(t)$, and therefore the
bound in (\ref{SuperAdiabaticThm}) basically yields (\ref{Aeps}).
Typically $\lim_{n\to\infty}C_n\eps^n= \infty$ for all $\eps>0$,
i.e.\ choosing $n$ larger while keeping $\eps$ fixed does not
necessarily decrease the bound in (\ref{SuperAdiabaticThm}).
However, one can choose $n_\eps=n(\eps)$ in such a way that
$C_{n_\eps}\eps^{n_\eps}$ is minimal. If $\theta$ is analytic, one
obtains the improved estimate
\[
\|\, P_- \, K^{n_\eps}_\eps(t,s)\, P_+ \| = \Or(\e^{-c/\eps})
\]
in the optimal superadiabatic basis $n_\eps$, see
\cite{Nen2,JoPf2}.

More interesting than bounds on $\mathcal{A}(\eps)$ is its actual
value. Since   $\mathcal{A}(\eps)$ is asymptotically smaller than
any power of $\eps$, this question is beyond standard perturbation
theory. For the case of analytic coupling $\theta$,  asymptotic
formulas of the type
\begin{equation}\label{LZ}
\mathcal{A}(\eps) = C\, \e^{-\frac{t_{\rm c}}{\eps}}\left(1 +
\Or(\eps)\right)
\end{equation}
have been established, see e.g.\ \cite{JKP,Jo}, where the
constants $C$ and $t_{\rm c}$ depend on the type and location of
the complex singularities of $\theta'(t)/\rho(t)$. However, these
results are obtained by solving (\ref{H}) not along the real axis
but along a Stoke's line in the complex plane. As a consequence
they give no information at all about the way in which the
exponentially small final transition amplitude $\mathcal{A}(\eps)$
is build up in real time. This question of adiabatic transition
histories is the concern of our paper. Berry \cite{Be} and, in a
refined way Berry and Lim \cite{BeLi,LiBe}, gave an  answer on a
non-rigorous level and explicitly left a mathematically rigorous
treatment as an interesting open problem. Only very recently
Hagedorn and Joye \cite{HaJo} succeeded and confirmed Berry's
results rigorously for a specific Hamiltonian.

Although our work has been strongly motivated by the findings of
Berry, our approach is slightly different. Let us first state our
main result before we discuss its relation to the earlier ones.
Without loss of generality we assume that $\rho(t)\equiv
\frac{1}{2}$. It was observed in \cite{Be}, that this can always
be achieved by transforming (\ref{H}) to the natural time scale
$$ \tau(t) = 2 \int_0^t \varrho(s) \, \D s\,.$$
However, we can only treat a rather special class of Hamiltonians,
since we must assume   that in the natural time scale the coupling
has the form
\begin{equation} \label{thetadot}
\theta'(t) = \I\gamma\left(\frac{1}{t+\I t_{\rm c}}-\frac{1}{t-\I
t_{\rm c}}\right)= \frac{\gamma t_{\rm c}}{t^2 + t_{\rm c}^2}
\end{equation}
with $\gamma\in\R$ and $t_{\rm c}>0$. In other words we assume
\begin{equation} \label{hamiltonian}
H(t) = \frac{1}{2} \left( \begin{array}{cc} \cos\theta(t) &\,\, \sin\theta(t) \\
\sin\theta(t) &\,\, - \cos\theta(t) \end{array}
\right)\qquad\mbox{with}\qquad \theta(t) =
2\,\gamma\arctan\left(\frac{t}{t_{\rm c}}\right)\,.
\end{equation}
We  shall comment below on the meaning of this special choice and
remark here that the Hamiltonian in \cite{HaJo} is
(\ref{hamiltonian}) with $\gamma=\frac{1}{2}$.

Our main result  is the construction of an optimal superadiabatic
basis in which the coupling term in the Hamiltonian is
exponentially small and can be computed explicitly at leading
order. This optimal basis is given as the $n_\eps^{\rm th}$
superadiabatic basis where $0\leq \sigma_\eps <2$ is such that
\begin{equation}\label{ndef}
n_\eps = \frac{t_{\rm c}}{\eps} -1+ \sigma_\eps\qquad\mbox{is an
even integer.}
\end{equation}

\begin{theorem} \label{MainThm} Let $H(t)$ be as in
(\ref{hamiltonian}) and
 $n_\eps$ as in (\ref{ndef}), and let  $\eps_0>0$  be sufficiently small. Then for every $\eps\in
(0,\eps_0]$ one can construct   a family of unitary matrices
$U_\eps^{n_\eps}(t)\in \C^{2\times 2}$, depending smoothly on
$t\in\R$, such that
\begin{equation}\label{UminusUnull}
\|U_\eps^{n_\eps}(t) - U_0(t)\|
=\Or\left(\frac{\eps^2}{1+t^2}\right)
\end{equation}
and
\begin{equation}\label{Hop}
U_\eps^{n_\eps}(t) \,\Big( \I \eps\partial_t -
H(t)\Big)\,U_\eps^{n_\eps *}(t) = \I\eps\partial_t -
\underbrace{\left(\begin{array}{cc} \rho_\eps^{n_\eps}(t)& c^{\op}_\eps(t) \\
\overline{c}^{\op}_\eps(t) & -\rho_\eps^{n_\eps}(t)
\end{array}\right)}_{\displaystyle =:H^{\op}_\eps(t)}\,.
\end{equation}
Here
\[
\rho_\eps^{n_\eps}(t)=\frac{1}{2} +
\Or\left(\frac{\eps^2}{1+t^2}\right)\,,
\]
and for every $\alpha<\frac{3}{2}$
\begin{equation}\label{Hod}
c^{\op}_\eps(t) = 2\I\,\sqrt{\frac{2\eps}{\pi t_{\rm
c}}}\,\sin\left(\frac{\pi \gamma}{2}\right)\, \e^{-\frac{t_{\rm
c}}{\eps}}\,\e^{-\frac{t^2}{2\eps t_{\rm c}}}
\,\cos\left(\frac{t}{\eps}-\frac{t^3}{3\eps t_{\rm c}^2} +
\frac{\sigma_\eps t}{t_{\rm c}} \right) +
\Or\left(\phi^\alpha(\eps,t)\right)\,,
\end{equation}
with
\begin{equation}\label{Phialpha}
\phi^\alpha(\eps,t) = \left\{ \begin{array}{ll}
        \eps^\alpha \exp\left(-\frac{t_{\rm c}}{\eps} \left( 1 +
        \frac{t^2}{4 t_{\rm c}^2} \right) \right)\ &\,\, \mbox{if }
         |t| < t_{\rm c},
        \\[3mm]
        {\displaystyle \frac{1}{1+t^2}} \exp \Big( -\frac{t_{\rm c}}{\eps}
        \Big(1+\frac{\ln 2}{2} \Big) \Big)  &\,\, \mbox{if }
         |t| \geq t_{\rm c}.
        \end{array} \right.
\end{equation}
\end{theorem}

\begin{remark}
 The explicit term in $c_\eps^{\op}$ is of order
$\Or(\e^{-t_{\rm c}/\eps})$ only for times $|t|=\Or(\sqrt{\eps})$.
For larger times all terms in $c_\eps^{\op}$ are exponentially
small compared to the leading exponential $\e^{-t_{\rm c}/\eps}$.
As a consequence, Taylor expansion of the cosine in $c_\eps^{\op}$
around $t/\eps$ for $|t|=\Or(\sqrt{\eps})$ shows that it can be
replaced by $\cos(t/\eps)$ at the cost of lowering $\alpha$ to
$\alpha<1$: for every $\alpha<1$
\[
c_\eps^{\op}(t) = 2\I\,\sqrt{\frac{2\eps}{\pi t_{\rm
c}}}\,\sin\left(\frac{\pi \gamma}{2}\right)\, \e^{-\frac{t_{\rm
c}}{\eps}}\,\e^{-\frac{t^2}{2\eps t_{\rm c}}}
\,\cos\left(\frac{t}{\eps} \right) +
\Or\left(\phi^\alpha(\eps,t)\right)\,.
\]
\end{remark}
\begin{remark}
The slow time decay of the error in (\ref{Phialpha}) for large
times is due to the fact that $n_\eps$ is optimal for $t$ near
$0$, but not for large $t$.
\end{remark}
\begin{remark}
  Taking $n_\eps$ defined in (\ref{ndef}) odd instead of even
  would yield slightly different off-diagonal elements in the effective
  Hamiltonian $H_\eps^{\op}(t)$. However, the resulting
  unitary propagator, cf.\ Corollary~\ref{PropCor}, would be the
  same at leading order. See the end of Section~5 for a discussion
  of this somewhat surprising fact.
\end{remark}

Let us shortly explain the idea of the proof of
Theorem~\ref{MainThm} and at the same time the structure of our
paper. First we construct the $n^{\rm th}$ order superadiabatic
basis as in (\ref{nthorderSAB}) in two steps: in Section~2 we
construct the projectors on the superadiabatic basis vectors and
in Section~3 we construct the unitary basis transformation
$U^n_\eps(t)$. We cannot use existing results here, e.g.\
\cite{Ga,Nen2}, since we need to keep carefully track of the exact
form off the off-diagonal terms $c^n_\eps(t)$ of the
superadiabatic Hamiltonian, and since we aim at a scalar
recurrence relation instead of a matrix recurrence relation for
the $c^n_\eps(t)$'s. The main mathematical challenge is the
asymptotic analysis of the resulting recurrence relation, which is
done in Section~4. This is also the only part where we have to
assume the special form (\ref{thetadot}) for $\theta'$.
Theorem~\ref{MainThm} then follows by choosing the order $n$ of
the superadiabatic basis as in (\ref{ndef}), a choice which
minimizes
 $c_\eps^{n}(t)$  near $t=0$. The details of this optimal truncation
 procedure and the proper proof of Theorem~\ref{MainThm} are
given in Section~5. Finally in Section~6 we use first order
perturbation theory in the optimal superadiabatic basis in order
to obtain the following Corollary, in which we abbreviate
\[
\Delta(t,s) := \arctan(t) -\arctan(s)\,.
\]
Also  recall that erf$:\R\to (-1,1)$ with ${\rm erf}(x) =
\frac{2}{\sqrt{\pi}} \int_0^x\e^{-x^2}\D x$ switches smoothly and
monotonically from erf$(-\infty)=-1$ to erf$(\infty)=1$.

\begin{corollary}\label{PropCor}
The unitary propagator in the optimal superadiabatic basis
\[
K_\eps^{\op}(t,s) = \left( \begin{array}{cc} k_\eps^+(t,s) &
k_\eps^{\od}(t,s)
\\\overline{k}_\eps^{\od}(t,s)& k_\eps^-(t,s)
\end{array}\right)\,,
\]
 i.e.\ the solution of
\[
\I\eps\partial_t K_\eps^{\op}(t,s) = H_\eps^{\op}(t)
K_\eps^{\op}(t,s) \,,\qquad K_\eps^{\op}(s,s) = {\rm id}\,,
\]
satisfies
\begin{equation}\label{Kdiag}
k_\eps^\pm(t,s) = \e^{\mp\frac{\I(t-s)}{2\eps}} +
\Or(\eps\Delta(t,s))
\end{equation}
and
\begin{eqnarray}\nonumber
k^{\od}_\eps(t,s) &=& \sin\left(\frac{\pi \gamma}{2}\right)\,
\e^{-\frac{t_{\rm c}}{\eps}}\e^{-\frac{\I(t+s)}{2\eps}}\left({\rm
erf}\left(\frac{t}{\sqrt{2\eps t_{\rm c}}}\right)-{\rm
erf}\left(\frac{s}{\sqrt{2\eps t_{\rm
c}}}\right)\right)\\&&+\,\Or\left(\sqrt{\eps}\e^{-\frac{t_{\rm
c}}{\eps}}\Delta(t,s)\right)\,.\label{Koffdiag}
\end{eqnarray}
Outside the transition region, more precisely  for
$|t|>\eps^\beta$ and $|s|>\eps^\beta$ for some
$\beta<\frac{1}{2}$, (\ref{Koffdiag}) holds with the error term
replaced by $\Or(\eps^\alpha\e^{-\frac{t_{\rm
c}}{\eps}}\Delta(t,s))$ for every $\alpha<1$.
\end{corollary}

Corollary~\ref{PropCor} immediately implies the existence of
solutions to (\ref{H}) of the form
\begin{equation}\label{approxsol}
\psi(t) = U_\eps^*(t)\,\left(\begin{array}{c}\e^{-\frac{\I
t}{2\eps}}\\\sin\left(\frac{\pi \gamma}{2}\right)\,
\e^{-\frac{t_{\rm c}}{\eps}}\e^{\frac{\I t}{2\eps}}\left({\rm
erf}\left(\frac{t}{\sqrt{2\eps t_{\rm c}}}\right)+1\right)
\end{array}\right)+\,\Or\left(\sqrt{\eps}\e^{-\frac{t_{\rm c}}{\eps}}\right)\,.
\end{equation}
They start at large negative times in the positive energy
adiabatic subspace and smoothly and monotonically develop the
exponentially small component in the negative energy adiabatic
subspace in a $\sqrt{\eps}$-neighborhood of $t=0$. Berry and Lim
\cite{Be,BeLi} argue that this behavior is universal: whenever
$\theta'$ has the form
\[
\theta'(t) = \frac{\pm\I\gamma}{t\pm\I t_{\rm c}}+ \Or(|t\pm\I
t_{\rm c}|^\alpha)\qquad\mbox{for some}\,\,\alpha>-1
\]
near its singularities $\pm\I t_{\rm c}$ closest to the real axis,
then (\ref{approxsol}) should hold. For the Landau-Zener
Hamiltonian, which describes the generic situation, one finds
$\gamma=\frac{1}{3}$ and $\alpha=-\frac{1}{3}$.
 Hagedorn and Joye \cite{HaJo}
proved (\ref{approxsol}) for the Hamiltonian (\ref{hamiltonian})
with $\gamma=\frac{1}{2}$. In the approach of Berry and, slightly
modified, of Hagedorn and Joye, the optimal superadiabatic basis
vectors are obtained through optimal truncation of  an asymptotic
expansion of the true solution of (\ref{H}) in powers of $\eps$.

In contrast, in our approach the optimal superadiabatic basis is
constructed by approximately diagonalizing the Hamiltonian.
 The main advantage of ``transforming the Hamiltonian''
over ``expanding the solutions'' is that the former approach can
be applied, at least heuristically, to more general adiabatic
problems, cf.\ \cite{Te}, as for example the Born-Oppenheimer
approximation. While we cannot control the asymptotics  for the
Born-Oppen\-heimer model rigorously yet,  the heuristic
application of the idea yields new physical insight into adiabatic
transition histories and new expressions for the exponentially
small off-diagonal elements of the $S$-matrix for simple
Born-Oppenheimer type models, cf.\ \cite{BeTe}. Therefore we see
the rigorous results obtained in this paper also as a first
attempt to justify  the application of analogous  ideas to more
complicated but also more relevant systems. Furthermore, the
concept of an adiabatically renormalized Hamiltonian was used to
derive a criterion for selecting possible transition sequences in
multi-level problems \cite{WiMo}.

 For the specific problem (\ref{H}) the
knowledge of two linearly independent solutions is of course
equivalent to the knowledge of the propagator and the effective
Hamiltonian in the optimal superadiabatic basis. Therefore we
shortly explain which aspects of our result constitute an
improvement compared to \cite{HaJo}: Most importantly, our proof
does not rely on the a priori knowledge of the scattering
amplitude $\mathcal{A}(\eps)$. Indeed, our result yields for the
first time a proof of (\ref{LZ}) based on superadiabatic
evolution, as expressed in Corollary~\ref{SmatrixCor}. Moreover,
 we allow for a slightly larger class of Hamiltonians and obtain more detailed
error estimates, which, in particular, give rise to close to
optimal error bounds in the expansion of the $S$-matrix, cf.\
Corollary~\ref{SmatrixCor}. Finally, we also get explicitly the
next order correction in (\ref{Koffdiag}) resp.\
(\ref{approxsol}), cf.\ Section~6. It should be noted, however,
that the improved error estimates and the next order corrections
could have been obtained also based on the proof in \cite{HaJo}.

 We finally turn to the scattering limit. Let $K_\eps^0(t,s)$ denote
the propagator in the original basis and define the scattering
matrix in the adiabatic basis by
\[
S_\eps^{\rm a} := \lim_{t\to\infty} \e^{\frac{\I H_0 t}{\eps}}\,
U_0(t)\,K_\eps^0(t,-t)\,U_0^*(-t)\, \e^{\frac{\I H_0 t}{\eps}}\,,
\quad\mbox{where}\,\, H_0 =
\left(\begin{array}{cc}\frac{1}{2}&0\\0&-\frac{1}{2}\end{array}\right)\,.
\]
Since, according to (\ref{UminusUnull}), for large negative and
positive times the optimal superadiabatic basis agrees with the
adiabatic basis, $S_\eps^{\rm a}$ can be computed with help of the
optimal superadiabatic propagator from Corollary~\ref{PropCor}.

\begin{corollary} \label{SmatrixCor}
For $\beta<1$ we have
\[
S_\eps^{\rm a} = \left(
\begin{array}{cc}
1 +\Or(\eps) & 2\,\sin\left(\frac{\pi \gamma}{2}\right)\,
\e^{-\frac{t_{\rm c}}{\eps}}\,(1+\Or(\eps^\beta))\\
2\,\sin\left(\frac{\pi \gamma}{2}\right)\, \e^{-\frac{t_{\rm
c}}{\eps}}\,(1+\Or(\eps^\beta))& 1+\Or(\eps)
\end{array}
\right)\,.
\]
\end{corollary}
\begin{proof}
According to (\ref{UminusUnull}) we have
\[
S_\eps^{\rm a} = \lim_{t\to\infty} \e^{\frac{\I H_0 t}{\eps}}\,
U_\eps^{n_\eps}(t)\,K_\eps^0(t,-t)\,U_\eps^{n_\eps\,\ast}(-t)\,
\e^{\frac{\I H_0 t}{\eps}} = \lim_{t\to\infty} \e^{\frac{\I H_0
t}{\eps}}\,K_\eps^{\op}(t,-t)\, \e^{\frac{\I H_0 t}{\eps}}\,.
\]
Now the claim follows  from inserting (\ref{Kdiag}) and
(\ref{Koffdiag}) with the improved error estimate outside of the
transition region. \qedi
\end{proof}

From Corollary~\ref{SmatrixCor} we conclude that the transition
amplitude is given by
\[
\mathcal{A}(\eps) = \|P_- S_\eps^{\rm a} P_+\| =
2\,\sin\left(\frac{\pi \gamma}{2}\right)\, \e^{-\frac{t_{\rm
c}}{\eps}}\,\left(1+\Or(\eps^\beta)\right)\,,\qquad\mbox{for any}
\,\,\beta<1\,,
\]
which agrees with the results of \cite{Jo}, as explained in
\cite{BeLi}.

We conclude the introduction with two recommendations for further
reading: The numerical results of Berry and Lim \cite{LiBe}
beautifully illustrate the idea of optimal superadiabatic bases
and universal adiabatic transition histories. The introduction of
the paper of Hagedorn and Joye \cite{HaJo} gives a slightly
different viewpoint on the problem and, in particular, a short
discussion on how exponential asymptotics for the Schr\"odinger
equation (\ref{H}) fit into the broader field of exponential
asymptotics for ordinary differential equations.

\medskip
\noindent{\bf Acknowledgements:}
 We are grateful to Alain Joye  and George Hagedorn for many helpful discussions.

\section{Superadiabatic projections}

For the present and the following section we assume that $H(t)$
has the form (\ref{hamiltonian}), but with some arbitrary
$\theta\in C^\infty(\R)$.
 The first aim is to construct time-dependent matrices $\pi^{(n)} \in \R^{2 \times 2}$ with
\begin{eqnarray}
 & & (\pi^{(n)})^2 - \pi^{(n)} = \Or(\varepsilon^{n+1}),  \label{pi1}\\
 & & \commut{\I\eps \partial_t - H}{\pi^{(n)}} = \Or(\varepsilon^{n+1}). \label{pi2}
\end{eqnarray}
Here, $\commut{A}{B} = AB - BA$ denotes the commutator two
operators $A$ and $B$. Likewise, we will later use $\antcom{A}{B}
= AB + BA$ to denote the anti-commutator of $A$ and $B$. Equation
(\ref{pi1}) says that $\pi^{(n)}$ is a projection up to errors of
order $\eps^{n+1}$, while (\ref{pi2}) implies that $\pi^{(n)}(t)$
is approximately equivariant, i.e.\
\[
K_\eps^0(t,s)\,\pi^{(n)}(s) = \pi^{(n)}(t)\,K_\eps^0(t,s) +
\Or(\eps^{n})\,.
\]
Recall the $K_\eps^0(t,s)$ is the unitary propagator for
(\ref{H}). Hence $\pi^{(n)}(t)$ is an almost projector onto an
almost equivariant subspace.

We construct $\pi^{(n)}$ inductively starting from the Ansatz
\begin{equation} \label{ansatz}
\pi^{(n)} = \sum_{k=0}^n \pi_k \varepsilon^k\,.
\end{equation}
By (\ref{hamiltonian}), $H$ has two eigenvalues $\pm1/2$. Let
$\pi_0$ be the projection onto the eigen\-space corresponding to
$+ 1/2$, and $\pi^{(0)} = \pi_0$ according to (\ref{ansatz}). It
is easily checked that (\ref{pi1}) and (\ref{pi2}) are fulfilled
for $n=0$. In order to construct $\pi_n$ for $n > 0$, let us write
$G_n(t)$ for the term of order $\eps^{n+1}$ in (\ref{pi2}), i.e.
\begin{equation} \label{G def}
(\pi^{(n)})^2 - \pi^{(n)} = \eps^{n+1} G_{n+1} +
\Or(\eps^{n+2})\,.
\end{equation}
Obviously,
\begin{equation} \label{G calculated}
G_{n+1} = \sum_{j=1}^{n} \pi_j \pi_{n+1-j}.
\end{equation}

\begin{proposition} \label{general scheme}
Assume that $\pi^{(n)}$ given by (\ref{ansatz}) fulfills
(\ref{pi1}) and (\ref{pi2}). Then a unique matrix $\pi_{n+1}$
exists such that $\pi^{(n+1)}$ defined as in (\ref{ansatz})
fulfills (\ref{pi1}) and (\ref{pi2}). $\pi_{n+1}$ is given by
\begin{equation}\label{recursion1}
\pi_{n+1} = G_{n+1} - \pi_0G_{n+1} - G_{n+1}\pi_0 - \I
\commut{\pi'_n}{\pi_0}.
\end{equation}
Furthermore $\pi'_n$ is off-diagonal with respect to $\pi_0$, i.e.
\begin{equation} \label{offdiag}
\pi_0  \pi'_n \pi_0  =  (1-\pi_0) \pi'_n (1-\pi_0) = 0,
\end{equation}
and $G_{n}$ is diagonal with respect to $\pi_0$, i.e.
\begin{equation} \label{diag}
\pi_0 G_{n+1} (1-\pi_0)  =  (1-\pi_0) G_{n+1} \pi_0 = 0.
\end{equation}
\end{proposition}
\begin{remark}
 The fact that the superadiabatic projections are unique
answers the question raised in \cite{Be} to which extend the
superadiabatic basis constructed there is uniquely determined.
\end{remark}
\begin{remark}
Our construction can be seen as a special case of the construction
in \cite{EmWe}. It was applied in the same context in
\cite{PST,Te}. The role and the importance of the superadiabatic
subspaces as opposed to the superadiabatic evolution have been
emphasized by Nenciu \cite{Nen2}. He constructs the superadiabatic
projections for much more general time-dependent Hamiltonians.
However, Nenciu's construction is less suitable for the explicit
computations we need to perform.
\end{remark}
\begin{proof}
Let $\pi^{(n+1)}$ be given by (\ref{ansatz}) and suppose
$\pi^{(n)}$ fulfills (\ref{pi1}) and (\ref{pi2}). Let
$\tilde\pi_{n+1}$ be an arbitrary matrix, and define
$\tilde\pi^{(n+1)} = \pi^{(n)} + \eps^{n+1} \tilde\pi_{n+1}$. Then
$$ (\tilde \pi^{(n+1)})^2 - \tilde \pi^{(n+1)} = (\pi^{(n)})^2 - \pi^{(n)} + \eps^{n+1}
\left( \antcom{\tilde \pi^{(n+1)}}{\tilde \pi_{n+1}} - \tilde \pi_{n+1} \right).$$
Using (\ref{G def}), we see that terms of order $\eps^{n+1}$
vanish if and only if
\begin{equation} \label{eq001}
G_{n+1} = \tilde \pi_{n+1} - \antcom{\pi_0}{\tilde \pi_{n+1}} =
(1-\pi_0) \tilde \pi_{n+1}(1-\pi_0) - \pi_0 \tilde \pi_{n+1}
\pi_0.
\end{equation}
Multiplying (\ref{eq001}) with $(1 - \pi_0)$ and with $\pi_0$ on
both sides and subtracting the results, we find that $\tilde
\pi_{n+1}$ must fulfill
\begin{equation} \label{eq002}
(1-\pi_0) \tilde \pi_{n+1}(1-\pi_0) + \pi_0 \tilde \pi_{n+1} \pi_0
= G_{n+1} - \antcom{G_{n+1}}{\pi_0}.
\end{equation}
Similarly
$$
 \commut{\I \eps \partial_t - H}{\tilde \pi^{(n+1)}}  =  \commut{\I \eps \partial_t - H}{\pi^{(n)}} + \eps^{n+1}
 \commut{\I \eps \partial_t - H}{\tilde \pi_{n+1}}.
 $$
 Again terms of order $\eps^{n+1}$ vanish if and only if
 \begin{equation} \label{offdiag explicit}
  \I \pi'_n = \commut{H}{\tilde \pi_{n+1}}.
\end{equation}
Since $\pi_0$ is the projector onto the eigenspace of $H$, we have
$\pi_0 H = H \pi_0 = E\pi_0$, where $E = 1/2$ is the positive
eigenvalue of $H$, and similarly $(1-\pi_0)H = H(1-\pi_0) =
-E(1-\pi_0)$. When we multiply (\ref{offdiag explicit}) first with
with $\pi_0$ from the left and with $1-\pi_0$ from the right, then
the other way round, and finally subtract the second result from
the first, we get
\begin{equation} \label{eq003}
2 E (\pi_0 \tilde \pi_{n+1} (1-\pi_0) + (1-\pi_0) \tilde \pi_{n+1}
\pi_0 ) = -\I \commut{\pi'_n}{\pi_0}.
\end{equation}
Now we divide (\ref{eq003}) by $2E$ and add (\ref{eq002}) to find
\begin{equation} \label{eq004}
\tilde \pi_{n+1} = G_{n+1} - \antcom{G_{n+1}}{\pi_0} -
\frac{\I}{2E} \commut{\pi'_n}{\pi_0}.
\end{equation}
Thus $\tilde\pi_{n+1}$ is uniquely determined by the requirement
that $\tilde\pi^{(n+1)}$ should fulfill (\ref{pi1}) and
(\ref{pi2}).  On the other hand, $\commut{H}{G_{n+1} -
\antcom{G_{n+1}}{\pi_0}} = 0$ and
\[\pi_0 \commut{\pi'_n}{\pi_0}
\pi_0 = (1-\pi_0) \commut{\pi'_n}{\pi_0} (1-\pi_0) = 0\,, \] and
thus $\pi_{n+1}$ given by the right hand side of (\ref{eq004})
indeed fulfills (\ref{eq002}) and (\ref{offdiag explicit}). This
shows existence. (\ref{diag}) and (\ref{offdiag}) now follow
directly from (\ref{eq001}) and (\ref{offdiag explicit}). \qedi
\end{proof}

The calculation of $\pi^{(n)}$ via the matrix recurrence relation
(\ref{recursion1}) and (\ref{G calculated}) is now possible in
principle, but extremely cumbersome. In order to make more
explicit calculations possible, we introduce a special basis of
$\R^{2 \times 2}$. Recall that $U_0(t)$ as defined in (\ref{U_0})
is the unitary transformation into the basis consisting of the
eigenvectors of $H$, i.e. the adiabatic basis, and let
 $V_0(t) = \frac{2}{\theta'(t)}U'_0(t)$. With $P=P_+$ as in (\ref{Pdef}) we then have $U_0^2 =
V_0^2 = \id$ and $PU_0V_0P = PV_0U_0P = 0$, and $\pi_0 = U_0PU_0$.
Moreover, since $G_1 = 0$ by (\ref{G calculated}),
(\ref{recursion1}) implies
\begin{equation} \label{pi_1}
\pi_1 = -\frac{\I}{2} \theta' (V_0PU_0 - U_0PV_0).
\end{equation}
Motivated by this, we put
\begin{eqnarray*}
X  =  V_0PU_0 -U_0PV_0\,, & \quad&
Y  =  V_0PV_0 - U_0PU_0\,, \\
Z  =  V_0PU_0 + U_0PV_0\,, & \quad & W  =  V_0PV_0 + U_0PU_0\,.
\end{eqnarray*}
It is immediate that this is a basis of $\R^{2 \times 2}$ for all
 $t$, and in fact
$$ X = \left(\begin{array}{cc}0 & -1 \\ 1 & 0 \end{array}\right), \quad W =
\left(\begin{array}{cc} 1 & 0 \\ 0 & 1 \end{array}\right), \quad Y = -2 H, \quad Z = \frac{-1}{\theta'}  Y'.$$
Our reason for representing $X$ through $Z$ via $U_0$ and $V_0$ is
that the following important relations now follow without effort:
\begin{eqnarray}
&&  X' = 0, \quad  Y' = -  \theta' Z, \quad  Z' =  \theta' Y, \label{derivatives} \\
&& \antcom XY = \antcom XZ = \antcom YZ = 0,\quad  -X^2 = Y^2 = Z^2 = W, \label{products} \\
&& \commut{X}{\pi_0} = Z, \quad \commut{Y}{\pi_0} = 0, \quad  \commut{Z}{\pi_0} = X, \label{commutators} \\
&& W - \antcom{W}{\pi_0} = Y. \label{W equation}
\end{eqnarray}
These relations show that this basis behaves extremely well under
the operations involved in the recursion (\ref{recursion1}). This
enables us to obtain

\begin{proposition} \label{function recursion}
For all $n \in \N$, $\pi_n$ is of the form
\begin{equation} \label{xyz ansatz}
\pi_n = x_nX + y_nY + z_nZ,
\end{equation}
where the functions $x_n, y_n $ and $z_n$ satisfy the differential
equations
\begin{eqnarray}
 x'_n & = &  \I  z_{n+1}, \label{E1a}\\
 y'_n & = & - \theta' z_n, \label{E1b}\\
z'_n & = &   \I  x_{n+1} + \theta'  y_n.  \label{E1c}
\end{eqnarray}
Moreover,
\begin{equation} \label{functions recursion start}
 x_1(t) = -\frac{\I}{2}\theta'(t), \quad y_1(t) = z_1(t) = 0.
\end{equation}
\end{proposition}

\begin{remark}
Hence, for all even $n$, $x_n = 0$, while for all odd $n$, $y_n =
z_n = 0$.
\end{remark}

\begin{proof}
(\ref{functions recursion start}) was already noticed in
(\ref{pi_1}). Now suppose $\pi_n$ is given by (\ref{xyz ansatz}).
By (\ref{products}) and (\ref{W equation}), $G_{n+1} -
\antcom{G_{n+1}}{\pi_0}$ is proportional to $Y$ with a prefactor
given through (\ref{G calculated}), and by (\ref{recursion1}),
(\ref{derivatives}) and (\ref{commutators}),
\begin{equation} \label{true recursion}
 \pi_{n+1} = \sum_{j=1}^n (- x_j x_{n+1-j} + y_j y_{n+1-j} + z_j z_{n+1-j}) Y + \I(\theta'y_n - z'_n) X - \I x'_nZ.
\end{equation}
Comparing with (\ref{xyz ansatz}) shows (\ref{E1a}) and
(\ref{E1c}). To show (\ref{E1b}), we use (\ref{offdiag}). This
gives
$$ 0 = \pi_0 \pi'_n \pi_0 = (y_n' + \theta' z_n) \pi_0Y\pi_0 + (z_n' - \theta'y_n)\pi_0Z\pi_0 + x'_n \pi_0X\pi_0.$$
Since $\pi_0Z\pi_0 = \pi_0X\pi_0 = 0$ and $\pi_0Y\pi_0 = \pi_0$,
the claim follows. \qedi
\end{proof}

\begin{remark}
From  (\ref{E1a}) through (\ref{E1c}) we may derive recursions for
calculating $x_n$ or $z_n$, e.g.
\begin{equation} \label{zn recursion with integration constant}
 z_{n+2}(t) = - \frac{\D}{\D t} \left( z'_n(t) +  \theta'(t)  \left( \int \theta'(t) z_n(t) \, \D t + C \right) \right).
\end{equation}
The constant of integration $C$ must (and in some cases can) be
determined by comparison with (\ref{true recursion}).
\end{remark}

Using (\ref{E1a})--(\ref{E1c}), we can give very simple
expressions for the quantities appearing in (\ref{pi1}) and
(\ref{pi2}). As for (\ref{pi2}), we use (\ref{derivatives}) and
the differential equations to find
\begin{eqnarray}
  \commut{\I\eps \partial_t - H}{\pi^{(n)}} & = & \I \eps^{n+1} \pi_n' = \I \eps^{n+1} \left(x_n' X + (y_n' + \theta' z_n) Y
  + (z_n' - \theta' y_n) Z\right) = \nonumber \\
  & = & - \eps^{n+1} (z_{n+1} X + x_{n+1} Z) \label{pi_n'}.
 \end{eqnarray}
Now we turn to $(\pi^{(n)})^2 - \pi^{(n)}$,  the term by which
$\pi^{(n)}$ fails to be a projector. Let us write
\begin{equation} \label{gnk}
(\pi^{(n)})^2 - \pi^{(n)} = \sum_{k=1}^{n} \eps^{n+k} G_{n+1,k}.
\end{equation}
With our earlier convention, $G_{n+1,1} = G_{n+1}$. Explicitly,
(\ref{ansatz}) and (\ref{gnk}) give
\begin{equation} \label{Gnk shape}
 G_{n+1,k} = \antcom{\pi_k}{\pi_n} + \antcom{\pi_{k+1}}{\pi_{n-1}} + \ldots =
 \sum_{j=0}^{n-k} \pi_{j+k} \pi_{n-j}.
 \end{equation}

\begin{proposition} \label{proj correction}
For each $n \in \N$, there exist functions $g_{n+1,k}, k \leq n$
with
\begin{equation} \label{projector error}
((\pi^{(n)})^2 - \pi^{(n)})(t) = \left(\sum_{k=1}^{n} \eps^{n+k}
g_{n+1,k}(t)\right) W.
\end{equation}
For each $k \leq n$,
$$g'_{n+1,k}  =  2 \I ( x_k  z _{n+1} -  z_k  x_{n+1}).$$
\end{proposition}
\begin{proof}
By (\ref{products}), each $G_{n+1,k}$ is proportional to $W$.
Using (\ref{xyz ansatz}) additionally, we find
$\antcom{\pi_k}{\pi_m} = 2 (- x_k  x_m +  y_k  y_m + 2  z_k
z_m)W$, and thus (\ref{Gnk shape}) yields
$$ g_{n+1,k} = \sum_{j=0}^{n-k} - x_{j+k}  x_{n-j} +  y_{j+k}  y_{n-j} + z_{j+k}  z_{n-j}.$$
Thus by using Proposition \ref{function recursion},
\begin{eqnarray*}
g'_{n+1,k} & = & \sum_{j=0}^{n-k} \I (  z_{j+k+1}  x_{n-j} +
x_{j+k}  z_{n-j+1} ) -
( \theta'  z_{j+k}  y_{n-j} +  \theta'  y_{j+k}  z_{n-j}) + \\
&& + \theta  y_{j+k}  z_{n-j} +   \theta'  y_{j+k}  z_{n-j} - \I (  x_{j+k+1}  z_{n-j} +  z_{j+k}  x_{n-j+1} ) = \\
& = & \I \sum_{j=0}^{n-k} ( (   z_{j+k+1}  x_{n-j} -   z_{j+k}  x_{n-j+1}) + ( x_{j+k}  z_{n-j+1} -   x_{j+k+1}  z_{n-j}) = \\
& = & 2 \I ( x_k z_{n+1} -  z_k  x_{n+1} ).
\end{eqnarray*}
The last equality follows because the sum is a telescopic sum.
\qedi
\end{proof}

Since $W = \id$ is independent of $t$, Proposition \ref{proj
correction} gives the derivative of the correction $(\pi^{(n)})^2
- \pi^{(n)}$ to a projector. As above, this gives an easy way for
estimating the correction itself provided we have some clue how to
choose the constant of integration.

\section{Construction of the unitary} \label{unitary}

We now proceed to construct the unitary transformation  $U_\eps^n$
into the $n^{\rm th}$ superadiabatic basis. By (\ref{ansatz}) and
(\ref{recursion1}), $\pi^{(n)}$ is self-adjoint. Thus it has two
orthonormal eigenvectors $v_n$ and $w_n$. Let
$$v_0 = \left( \begin{array}{c} \cos(\theta/2) \\ \sin(\theta/2) \end{array} \right), \quad
w_0 = \left( \begin{array}{c} \sin(\theta/2) \\ -\cos(\theta/2)
\end{array} \right)$$ be the eigenvectors of $\pi_0$, and write
\begin{equation} \label{v_n ansatz}
 v_n = \alpha v_0 + \beta w_0, \quad w_n = \overline{\alpha} w_0 - \overline{\beta} v_0 \qquad (\alpha, \beta \in \C).
 \end{equation}
We make this representation unique by requiring $0 \leq \alpha \in
\R$. Let $U_\eps^n$ be the unitary operator taking  $(v_n,w_n)$ to
the standard basis $(e_1,e_2)$ of $\R^2$ , i.e.
\begin{equation} \label{U}
  U_\eps^n = e_1 v_n^{\ast} + e_2 w_n^{\ast},
 \end{equation}
where all vectors are column vectors. Note that the definition
(\ref{U_0}) of $U_0$ is consistent with (\ref{U}) for $n=0$.
$U_\eps^n$ diagonalizes $\pi^{(n)}$, thus
\begin{equation} \label{diag pi_n}
  U_\eps^n \pi^{(n)} U_\eps^{n\, \ast} = D \equiv \left( \begin{array}{cc} \lambda_1 & 0 \\
        0 & \lambda_2 \end{array} \right),
\end{equation}
where $\lambda_{1,2}$ are the eigenvalues of $\pi^{(n)}$. Although
$\alpha, \beta$ and $\lambda_{1,2}$ depend on $n$, $\eps$ and $t$,
we suppress this from the notation.

\begin{lemma} \label{U errors}
$$ U_0 U_\eps^{n\,\ast} = \left( \begin{array}{cc} \alpha & -\overline{\beta} \\
                                \beta & \alpha \end{array} \right),  \quad
    \mbox{and} \quad U_0 U_\eps^{n\,\ast '} =
    \left( \begin{array}{cc} \alpha' + \beta & \,\,\alpha -\overline{\beta}' \\
                    \beta' - \alpha & \,\,\alpha' + \overline{\beta} \end{array} \right). $$
\end{lemma}
\begin{proof}
The calculations are straightforward and we only show the second
equality. First note that $v_0' = - w_0$ and $w_0' = v_0$. Thus
$$ U_\eps^{n\,\ast '} = ((\alpha' + \beta) v_0 + (\beta' - \alpha) w_0) e_1^{\ast} +
(( \alpha - \overline{\beta}')v_0 + (\alpha' +
\overline{\beta})w_0)e_2^\ast,$$ and using the orthogonality of
$v_0$ and $w_0$ yields the claim,
$$ U_0 U_\eps^{n\,\ast '} = e_1 (\alpha' + \beta) e_1^\ast + e_1 ( \alpha -
\overline{\beta}') e_2^\ast + e_2 (\beta' - \alpha) e_1^\ast + e_2
( \alpha' +\overline{\beta})e_2^{\ast}\,.\qquad\qedi$$
\end{proof}

It will turn out that $\beta, \alpha' \alpha$, and $\beta'$  are
small  quantities, $\lambda'_1, \lambda'_2$, and $\lambda_2$ are
even much smaller, while $\alpha^2$ and $\lambda_1$ are large,
i.e. of order $1 + \Or(\eps)$. This motivates the form in which we
present the following result.

\begin{proposition} \label{general transformed matrix}
Suppose $\lambda_1 \neq \lambda_2$. Then for each $n \in \N$,
$$ U_\eps^n (\I \eps \partial_t - H) U_\eps^{n\,\ast} =  \I \eps \partial_t -
\left( \begin{array}{cc}  \frac{1}{2} &  \frac{\alpha^2 \eps^{n+1}}{\lambda_1 - \lambda_2} (x_{n+1} - z_{n+1}) \\
        \frac{\alpha^2 \eps^{n+1}}{\lambda_1 -\lambda_2} (-x_{n+1} - z_{n+1}) & -\frac{1}{2} \end{array} \right) + R,$$
with
$$ R = {\footnotesize\left( \begin{array}{cc}   \eps \tIm (\overline{\beta}(2 \alpha + \beta'))
 + |\beta|^2 &
- \frac{\eps^{n+1} \overline{\beta}^2}{\lambda_1 - \lambda_2} (x_{n+1} + z_{n+1})  \\
        \frac{\eps^{n+1} \beta^2}{\lambda_1 -\lambda_2} (x_{n+1} - z_{n+1})  &
         - \eps \tIm (\overline{\beta}(2 \alpha + \beta')) - |\beta|^2   \end{array} \right)}.$$
\end{proposition}
\begin{proof}
Let us write $U_\eps^n (\I \eps \partial_t - H) U_\eps^{n\,\ast} =
(M_{i,j})$, $i,j \in \{1,2\}$. $M_{1,1}$ and $M_{2,2}$ are
calculated in a straightforward manner, using Lemma \ref{U errors}
together with the
 fact $U_0 H U_0^{\ast} = {\scriptsize \left( \begin{array}{cc} 1/2 & 0 \\
                    0 & -1/2 \end{array} \right)}$:
\begin{eqnarray*}
\lefteqn{U_\eps^n (\I \eps \partial_t - H) U_\eps^{n\,\ast}  =  \I
\eps
\partial_t + \I \eps U_\eps^n U_0^{\ast} U_0 U_\eps^{n\,\ast '} -
U_\eps^n U_0^{\ast} U_0 H U_0^{\ast} U_0 U_\eps^{n\,\ast} = }\\
&=& \I \eps \partial_t + \I \eps {\footnotesize
    \left( \begin{array}{cc} \alpha & \overline{\beta} \\
                                -\beta & \alpha \end{array} \right)
     \left( \begin{array}{cc} \alpha' + \beta & \alpha -\overline{\beta}' \\
            \beta' - \alpha & \alpha' + \overline{\beta} \end{array} \right) -
    \frac{1}{2} \left( \begin{array}{cc} \alpha & \overline{\beta} \\
                -\beta & \alpha \end{array} \right)
    \left( \begin{array}{cc} 1 & 0 \\
        0 & -1 \end{array} \right)
    \left( \begin{array}{cc} \alpha & -\overline{\beta} \\
                                \beta & \alpha \end{array} \right)}.
\end{eqnarray*}
Carrying out the matrix multiplication yields
\begin{equation} \label{the diag elements}
 M_{1,1} = - M_{2,2} = \I \eps \partial_t + \I \eps ((\alpha(\alpha' + \beta) +
 \overline{\beta}(\beta' - \alpha)) - \frac{1}{2} (\alpha^2 - |\beta|^2).
 \end{equation}
We now use $\alpha^2 + |\beta|^2 = 1$ to obtain $ 0 = 2 \alpha
\alpha' + \beta' \overline{\beta} + \overline{\beta}' \beta = 2
\tRe(\alpha \alpha' + \overline{\beta}\beta')$ and $\alpha^2
-|\beta|^2 = 1 - 2 |\beta|^2$. Plugging these into (\ref{the diag
elements})
 gives the diagonal coefficients of $M$.
 Although we could get expressions for the off-diagonal
coefficients by the same method, these would not be useful later
on. Instead we use (\ref{diag pi_n}), i.e. $U_\eps^{n\,\ast} D =
\pi^{(n)} U_\eps^{n\,\ast}$ together with (\ref{pi_n'}) and obtain
\begin{equation} \label{commut trick}
 U_\eps^n (\I \eps \partial_t - H) U_\eps^{n\,\ast}D = DU_\eps^n (\I \eps \partial_t - H)
 U_\eps^{n\,\ast} - \eps^{n+1} U_\eps^n (z_{n+1} X + x_{n+1} Z) U_\eps^{n\,\ast}.
\end{equation}
By multiplying (\ref{commut trick}) with $e_j e_j^{\ast}$ from the
left and by $e_k e_k^{\ast}$ from the right $(j,k \in \{1,2\})$
and rearranging, we obtain
\begin{eqnarray} \label{commut trick 2}
\lefteqn{(\lambda_k - \lambda_j) \,e_j \,e_j^{\ast} \,U_\eps^n (\I
\eps
\partial_t - H)\,
U_\eps^{n\,\ast} e_k \,e_k^{\ast} = } \\
& = & - \eps^{n+1}
 e_j \,e_j^{\ast}  \,U_\eps^n (z_{n+1} X + x_{n+1} Z) \,U_\eps^{n\,\ast} e_k\,
 e_k^{\ast} - \I \delta_{k,j} \,\eps \,\lambda'_j \,e_j\, e^{\ast}_j\,. \nonumber
 \end{eqnarray}
From the equalities
$ U_0 X U_0^{\ast} = { \left( \begin{array}{cc} 0 & 1 \\
                    -1 & 0 \end{array} \right)}$, $U_0 Z U_0^{\ast} =
                    { \left( \begin{array}{cc} 0 & -1 \\
                    -1 & 0 \end{array} \right)}$
and Lemma \ref{U errors} we obtain
$$ U_\eps^n X U_\eps^{n\,\ast} = {\footnotesize \left( \begin{array}{cc}
\alpha(\beta-\overline\beta) & \alpha^2 + \overline\beta^2 \\
        -(\alpha^2+\beta^2) & -\alpha(\beta - \overline{\beta}) \end{array} \right)}, \qquad
        U_\eps^n Z U_\eps^{n\,\ast} = {\footnotesize \left( \begin{array}{cc}
        -\alpha(\beta+\overline\beta) & -(\alpha^2 - \overline\beta^2) \\
        -(\alpha^2-\beta^2) & \alpha(\beta + \overline{\beta}) \end{array} \right)}.$$
The expressions for $M_{1,2}$ and $M_{2,1}$ follow by taking $k
\neq j$ in (\ref{commut trick 2}). \qedi \end{proof}

We now use our results from the previous section to express
$\alpha, \beta$ and $\lambda_{1,2}$ in terms of $x_k, y_k$ and
$z_k$, $k \leq n$. Let us define
\begin{eqnarray}
\label{xi} \xi & \equiv & \xi(n,\eps,t) = {\textstyle \sum_{k=1}^n} \eps^k x_k(t), \\
\label{eta} \eta & \equiv & \eta(n,\eps,t) = {\textstyle\sum_{k=1}^n} \eps^k y_k(t), \\
\label{zeta} \zeta & \equiv & \zeta(n,\eps,t) =
{\textstyle\sum_{k=1}^n} \eps^k z_k(t).
\end{eqnarray}
Moreover, let
\begin{equation} \label{g}
 g \equiv g(n,\eps,t) = {\textstyle\sum_{k=1}^{n}} \eps^{n+k} g_{n+1,k}(t)
\end{equation}
be the quantity appearing in (\ref{projector error}).

\begin{lemma} \label{eigenvalues}
The eigenvalues of $\pi^{(n)}$ solve the quadratic equation
$$ \lambda_{1,2}^2 - \lambda_{1,2} - g = 0.$$
\end{lemma}
\begin{proof}
By (\ref{diag pi_n}) and Proposition \ref{proj correction} we
obtain
$$ \left(\begin{array}{cc} \lambda_1^2 - \lambda_1 & 0 \\ 0 & \lambda_2^2 -
\lambda_2 \end{array} \right) = U_\eps^n ((\pi^{(n)})^2 -
\pi^{(n)}) \,U_\eps^{n\,\ast} = U_\eps^n g W U_\eps^{n\,\ast} =
\left(
\begin{array}{cc} \,g\, & \,0\, \\[2pt] \,0\, & \,g\, \end{array} \right)\,.\qquad\qedi $$
\end{proof}

\begin{lemma} \label{alpha beta}
$$ \alpha^2 (\lambda_1 - \lambda_2) = 1 - \eta - \lambda_2, \qquad \mbox{and }
\qquad \alpha \beta (\lambda_1 - \lambda_2)  = - \xi - \zeta.$$
\end{lemma}
\begin{proof}
We have
\begin{equation} \label{no1}
 \pi^{(n)} = \lambda_1 v_n v_n^{\ast} + \lambda_2 w_n w_n^{\ast}.
\end{equation}
Plugging in (\ref{v_n ansatz}), we obtain
\begin{eqnarray*}
\pi^{(n)} v_0 &=& \lambda_1 \alpha v_n - \lambda_2 \beta w_n = (\lambda_1\alpha^2 +
\lambda_2 |\beta|^2) v_0 + (\lambda_1 - \lambda_2)\alpha \beta w_0 = \\
&=& (\alpha^2(\lambda_1 - \lambda_2) + \lambda_2) v_0 + (\lambda_1
- \lambda_2)\alpha \beta w_0.
\end{eqnarray*}
In the last step, we used $|\beta|^2 + \alpha^2 = 1$. On the other
hand, from (\ref{ansatz}) and (\ref{recursion1}) we have
\begin{equation}\label{p0minuspi}
\pi^{(n)} = \pi_0 + \sum_{k=1}^n \eps^k (x_k X + y_kY + z_kZ)\,,
\end{equation}
and since $Xv_0 = Zv_0 = -w_0$, $\pi_0 v_0 = v_0$ and $Y v_0 = -
v_0$, we find
$$ \pi^{(n)} v_0 = (1 - \eta) v_0 - (\xi + \zeta) w_0.$$
Comparing coefficients finishes the proof. \qedi \end{proof}

\begin{theorem} \label{general diag} Let $\eps_0>0$ be
sufficiently small. For $\eps\in(0,\eps_0]$ assume there is a
bounded function $q$ on $\R$ such that $\xi(t)$, $\eta(t)$,
$\zeta(t)$ and their derivatives $\xi'(t), \eta'(t), \zeta'(t)$
are all bounded in norm by $\eps q(t)$. Then
\begin{eqnarray} \label{general offdiag eq}
  \lefteqn{U_\eps^n
 (\I \eps \partial_t - H) U_\eps^{n\,\ast} =}\\&=& \I \eps \partial_t-
\left( \begin{array}{cc} \frac{1}{2} + \Or(\eps^2 q^2) &
\eps^{n+1} (x_{n+1} - z_{n+1})\, (1 + \Or(\eps q))
\\[2mm]
\eps^{n+1} (-x_{n+1} - z_{n+1})\, (1 + \Or(\eps q)) & -\frac{1}{2}
+ \Or(\eps^2 q^2) \end{array} \right)\,.\nonumber
\end{eqnarray}
\end{theorem}
\begin{proof}
From (\ref{p0minuspi}) and our assumptions it follows that
$\pi^{(n)} - \pi_0 = \Or(\eps q)$. Thus $\lambda_1 = 1 + \Or(\eps
q)$ and $\lambda_2 = \Or(\eps q)$, and from Lemma
\ref{eigenvalues} we infer $g = \Or(\eps)$ and
$$ \lambda_1 = {\textstyle\frac{1}{2}} \left( 1 + \sqrt{1 + 4 g} \right)\,,\qquad\lambda_2
= {\textstyle\frac{1}{2}} \left( 1 - \sqrt{1 + 4 g} \right)\,.$$
Since $\lambda_1 - \lambda_2 \neq 0$, Lemma \ref{alpha beta}
yields
$$ \alpha^2 = \frac{1 + \sqrt{1+4g} - 2 \eta}{2 \sqrt{1 + 4g}}, \quad \beta =
\frac{-\xi - \zeta}{\sqrt{1+4g} \alpha}\,.$$ Hence $ \alpha^2 = 1
+ \Or(\eps q),$ and  $\beta, \beta' $ and $ \alpha \alpha' =
(\alpha^2)' /2$ are all $\Or(\eps q)$. Plugging these into the
matrix $R$ in Proposition \ref{general transformed matrix} shows
the claim. \qedi
\end{proof}

\section{Solving the recursion: a pair of simple poles}

In order to make further progress, we need to understand the
asymptotic behavior of the off-diagonal elements of the effective
Hamiltonian in the $n^{\rm th}$ superadiabatic basis for large
$n$. According to (\ref{general offdiag eq}) this amounts to the
asymptotics of $x_n$ and $z_n$ as given by the recursion from
Proposition \ref{function recursion}. It is clear that the
function $\theta'$ alone determines the behavior of this
recursion. We will study here the special case
\begin{equation} \label{theta ansatz}
\theta'(t) = \frac{\I \gamma}{t+\I t_{\rm c}} - \frac{ \I
\gamma}{t-\I t_{\rm c}} = \frac{\gamma t_{\rm c}}{t^2 + t_{\rm
c}^2}\,.
\end{equation}
The reason lies in the intuition that the poles of $\theta'$
closest to the real axis determine the superadiabatic transitions,
and that these transitions are of universal form whenever these
poles are of order one, see \cite{Be,BeLi} for details. As in
\cite{HaJo}, we have to restrict to the special case that
$\theta'$ has no contribution besides these poles in order solve
the recursion. We now have two parameters left in $\theta'$. The
distance $t_{\rm c}$ of the poles from the real axis determines
the exponential decay rate in the in the off-diagonal elements of
the Hamiltonian and the strength of the residue $\gamma$
determines the pre-factor in front of the exponential.
 As is done in \cite{HaJo}, we could get rid of the parameter $t_{\rm c}$
by rescaling time, but we choose not to do so because $t_{\rm c}$
plays a nontrivial role in optimal truncation and the error bounds
obtained therein, and keeping this parameter will make things more
transparent.

We use (\ref{zn recursion with integration constant}) in order to
determine the asymptotics of $z_n$. From Proposition \ref{function
recursion} together with (\ref{true recursion}) it is clear that
$y_n$ must go to zero as $t \to \pm \infty$. This fixes the
constant of integration in (\ref{zn recursion with integration
constant}), and we arrive at the linear two-step recursion
\begin{equation} \label{zn recursion}
 z_{n+2}(t) = - \frac{\D}{\D t} \left( z'_n(t) +  \theta'(t) \int_{-\infty}^t \theta'(s) z_n(s) \, \D s  \right).
\end{equation}
The fact that the recursion is linear will make its analysis
simpler than the one of the nonlinear recursion in \cite{HaJo}.
 We rewrite $\theta'$ as
$$ \theta'(t) = \frac{\gamma}{t_{\rm c}} (f + \overline f) \qquad \mbox{with}\qquad f(t)
= \frac{\I t_{\rm c}}{t + \I t_{\rm c}}\,.$$
For $z_n$, we will make an Ansatz as a sum of powers of $f$ and
$\overline{f}$. The reason for the success of this approach is
the fact that this representation is stable under differentiation
and integration, and also under multiplication with $\theta'$
through the partial fraction expansion. More explicitly, we have

\begin{lemma} \label{identities}
For each $m \geq 1$,
\begin{eqnarray}
\theta' \tIm(f^m) &=& \frac{\gamma}{t_{\rm c}} \sum_{k=0}^{m-1} 2^{-k} \tIm(f^{m+1-k}), \label{Im mult} \\
\theta' \tRe(f^m) &=& \frac{\gamma}{t_{\rm c}} \left( \sum_{k=0}^{m-1} 2^{-k}
 \tRe(f^{m+1-k}) +  2^{-m} \theta' \right), \label{Re mult} \\
\tIm(f^m)' &=& -\frac{m}{t_{\rm c}} \tRe(f^{m+1}), \label{Im diff}\\
\tRe(f^m)' &=& \frac{m}{t_{\rm c}} \tIm(f^{m+1}). \label{Re diff}
\end{eqnarray}
\end{lemma}
\begin{proof}
We have $ f + \overline{f} = \frac{2t_{\rm c}^2}{t^2+t_{\rm c}^2}
= 2 f\overline{f}$, and thus
$$ f^k \overline f = \frac{1}{2} f^{k-1}(f+\overline f) = \frac{1}{2} (f^k + f^{k-1}\overline f)$$
and
\begin{equation} \label{multiplication rule}
\theta' f^{n-j} = \frac{\gamma}{t_{\rm c}} \left( \sum_{k=0}^{n-j}
2^{-k} f^{n+1 - (j+k)} + 2^{-n+j}\overline f \right).
\end{equation}
Taking the complex conjugate of (\ref{multiplication rule}) and
adding it to resp.\ subtracting it from  (\ref{multiplication
rule}), we arrive at (\ref{Im mult}) and (\ref{Re mult}). To prove
(\ref{Im diff}) and (\ref{Re diff}), it suffices to use that
$(f^k)' = k f^{k+1} / (\I t_{\rm c})$ along with the complex
conjugate equation. \qedi
\end{proof}


\begin{proposition} \label{z_n and recursion}
For each even $n \in \N$ and $j = 0, \ldots, n-1$, let the numbers
$a_j^{(n)}$ be recursively defined through
\begin{eqnarray}
a_0^{(2)} & = & 1, \qquad a_1^{(2)} = 0\,, \label{a_n rekursion start}\\
a_j^{(n+2)} &=& \frac{n+1-j}{(n+1)\,n} \left( (n-j)\,a_j^{(n)} -
\gamma^2 \sum_{k=0}^j \frac{1}{n-k}
\sum_{m=0}^k a_m^{(n)} \right) \quad (j<n)\,, \label{a_n rekursion step} \\
a_n^{(n+2)} & = & a_{n-1}^{(n+2)}, \qquad a_{n+1}^{(n+2)} = 0\,.
\nonumber
\end{eqnarray}
Then
\begin{eqnarray}
 z_n & = & - \gamma \frac{(n-1)!}{t_{\rm c}^n} \sum_{j=0}^{n-1} 2^{-j} a_j^{(n)}
 \tIm(f^{n-j}) \quad (n \mbox{ even})\,,
  \label{z_n recursion step} \\
 y_n & = &  \gamma^2 \frac{(n-1)!}{t_{\rm c}^{n}} \sum_{j=0}^{n-1} 2^{-j} \left(
 \frac{1}{n-j} \sum_{k=0}^j a_k^{(n)}\right) \tRe(f^{n-j})
                    \quad (n \mbox{ even})\,,  \label{y_n recursion step} \\
 x_n & = &  \I \gamma \frac{(n-1)!}{t_{\rm c}^{n}} \sum_{j=0}^{n-1} 2^{-j} \left(
 \frac{n}{n-j} a_j^{(n+1)}
 \right) \tRe(f^{n-j}) \quad (n \mbox{ odd})\,,  \label{x_n recursion step}
\end{eqnarray}
\end{proposition}
\begin{proof}
We proceed by induction. We have $  x_1 = \I  \theta' / 2 =
\frac{\I \gamma}{t_{\rm c}} \tRe(f)$, and thus by (\ref{E1a}) and
(\ref{Re diff}),
$$   z_2 =  \frac{\I}{t_{\rm c}}   x'_1 = - \frac{\gamma}{t_{\rm c}^2} \tIm(f^2).$$
This proves (\ref{z_n recursion step}) for $n=2$. Now suppose that
(\ref{z_n recursion step}) holds for some even  $n \in \N$. Then
by (\ref{E1a}) and (\ref{Re diff}), (\ref{x_n recursion step})
holds for $n-1$. To prove (\ref{y_n recursion step}) for the given
$n$, we want to use (\ref{E1b}). (\ref{Im mult}) and the induction
hypothesis on $z_n$ yield
\begin{eqnarray}
  \theta'   z_n & = & - \gamma^2 \frac{(n-1)!}{t_{\rm c}^{n+1}} \sum_{j=0}^{n-1}a_j^{(n)}
  \sum_{k=0}^{n-j-1} 2^{-(k+j)} \tIm(f^{n+1-(j+k)})  = \nonumber \\
& = &  - \gamma^2 \frac{(n-1)!}{t_{\rm c}^{n+1}} \sum_{m=0}^{n-1}
2^{-m} \left(\sum_{j=0}^m a_j^{(n)}\right) \tIm(f^{n+1-m})
\label{E5}.
\end{eqnarray}
Since (\ref{E5}) only contains second or higher order powers of
$f$, it is easy to integrate using (\ref{Re diff}). Let us write
\begin{equation} \label{bj in proof}
b_m^{(n)} = \frac{1}{n-m} \sum_{j=0}^m a_j^{(n)}.
\end{equation}
Then by (\ref{Re diff}) we obtain
$$  y_n = - \int_{-\infty}^t  \theta'(s)   z_n(s) \, \D s =  r^2 \frac{(n-1)!}{t_{\rm c}^{n-1}}
\sum_{m=0}^{n-1} 2^{-m} b_m^{(n)} \tRe(f^{n-m}),$$ proving
(\ref{y_n recursion step}) for $n$. It remains to prove (\ref{z_n
recursion step}) for $n+2$. We want to
 use (\ref{zn recursion}), and therefore we employ (\ref{Re mult}) and our above calculations in order to get
\begin{eqnarray*}
 \lefteqn{ \theta'(t ) \int_{-\infty}^t    \theta'(s)   z_n(s) \, \D s
 = }\\&=&  - \gamma^3 \frac{(n-1)!}{t_{\rm c}^{n+1}}
\sum_{j=0}^{n-1} b_j \left( \sum_{k=0}^{n-j+1} 2^{-(k+j)} \tRe(f^{n+1-(k+j)}) +  2^{-n}   \theta' \right) =   \\
& =&  - \gamma^3 \frac{(n-1)!}{t_{\rm c}^{n+1}} \left( \left(
\sum_{j=0}^{n-1} 2^{-j} \left( \sum_{k=0}^j b_k \right)
\tRe(f^{n+1-j})\right)  + 2^{-n+1}  \left( \sum_{k=0}^{n-1} b_k
\right) \tRe(f) \right).
\end{eqnarray*}
By (\ref{Im diff}),
$$    z'_n = \gamma \frac{(n-1)!}{t_{\rm c}^{n+1}} \sum_{j=0}^{n-1} 2^{-j} a_j^{(n)} (n-j) \tRe(f^{n+1-j}).$$
Now we sum the last two expressions, differentiate again and
obtain
\begin{eqnarray*}
  z_{n+2} &=&  - \gamma \frac{(n-1)!}{t_{\rm c}^{n+2}} \left( \sum_{j=0}^{n-1} 2^{-j} (n+1-j)
   \left( (n-j) a_j^{(n)} - \gamma^2 \sum_{k=0}^j b_k \right) \tIm(f^{n+2-j}) \right. \\
&& \left.  - 2 \gamma^2 2^{-n} \left(\sum_{k=0}^{n-1} b_k\right)
\tIm(f^2)  \right).
\end{eqnarray*}
Comparing coefficients, this proves (\ref{z_n recursion step}) for
$n+2$. \qedi
\end{proof}

We now investigate the behavior of the coefficients $a_j^{(n)}$ as
$n \to \infty$.

\begin{proposition} \label{a_n limit}
Let $a_j^{(n)}$ be defined as in Proposition~\ref{z_n and
recursion}. \alp
\begin{enumerate}
\item  $a_0^{(n)} = {\displaystyle \frac{\sin (\gamma\pi /
2)}{\gamma \pi/2} }\left(1 + \Or\left(\frac{\gamma^2}{n^2}\right)\right).$\\
\item There exists $C_1 > 0$ such that for all $n \in \N$
$$ |a_1^{(n)}| \leq C_1 \frac{\ln n}{n-1}\,.$$
 \item For each $p>1$ there exists $C_2 > 0$
such that for all $n \in \N$
$$ \sup_{j \geq 2} p^{-j} |a_j^{(n)}| \leq \frac{C_2}{n-1}\,.$$

\end{enumerate}
\end{proposition}

\begin{proof}
(a) By (\ref{a_n rekursion start}), $a_0^{(2)} = 1$, and
$$ a_0^{(n+2)}  = a_0^{(n)}\left( 1 - \frac{\gamma^2}{n^2} \right).$$
Comparing with the product representation of the sine function,
$$ \sin(\pi x) = \pi x \prod_{n=1}^{\infty} \left( 1 - \frac{x^2}{n^2} \right)\,,$$
we arrive at (a). \\
(b) Put $\alpha_n = (n-1) a_1^{(n)}$. Then by (\ref{a_n rekursion
step}),
$$ \alpha_{n+2} = \alpha_n \left( 1 - \frac{\gamma^2}{(n-1)^2} \right) - \gamma^2
\left( \frac{1}{n} + \frac{1}{n-1} \right) a_0^{(n)}.$$ thus for
$n-1 > \gamma$, we have
$$ |\alpha_{n+2}| \leq |\alpha_n| + \gamma^2 \left( \frac{1}{n} + \frac{1}{n-1} \right)
\max_{m \in \N} |a_0^{(m)}|,$$
which shows (b). \\
(c) Put $c_j^{(n)} = (n-1) p^{-j} a_j^{(n)}$, and $c^{(n)} =
\max_{j \geq 2} |c_j^{(n)}|$. We will show that the sequence
$c^{(n)}$ is bounded. We have
\begin{eqnarray}
c_j^{(n+2)} & = & \frac{n+1-j}{n(n-1)} \left( (n-j) c_j^{(n)} - \gamma^2 \sum_{k=2}^j \frac{1}{n-k}
\sum_{m=2}^k p^{-j+m} c_m^{(n)} - \right. \nonumber \\
&& \left. -(n-1) p^{-j} \gamma^2 \left( a_0^{(n)} \sum_{k=0}^j
\frac{1}{n-k} + a_1^{(n)} \sum_{k=1}^j \frac{1}{n-k} \right)
\right). \label{bigformula}
\end{eqnarray}
Now
\begin{equation} \label{term1}
 \left| \sum_{k=2}^j \frac{1}{n-k} \sum_{m=2}^k p^{-j+m} c_m^{(n)} \right|
\leq c^{(n)} \frac{1}{n-j}\frac{p^2}{(p-1)^2},
\end{equation}
and
\begin{equation} \label{term2}
 p^{-j} \sum_{k=0}^j \frac{1}{n-k} \leq \frac{(j+1) p^{-j}}{n-j} \leq \frac{1}{(n-j)\ln p}.
\end{equation}
We plug these results into (\ref{bigformula}) and obtain
\begin{eqnarray*} 
|c_j^{(n+2)}| &\leq& c^{(n)} \left( \frac{(n+1-j)(n-j)}{n(n-1)}  +
\frac{(n+1-j)\,\gamma^2 p^2}{(n-j)(p-1)^2} \frac{1}{n(n-1)} \right) +\\
& & + \frac{1}{n} \frac{(n+1-j)\,p^2\,\gamma^2}{(n-j) (p-1)^2 \ln
p}(|a_0^{(n)}|+|a_1^{(n)}|).
\end{eqnarray*}
By (a) and (b),  $a_0^{(n)}$ and $a_1^{(n)}$ are bounded. Taking
the supremum over $j \geq 2$ above, we see that  there exist
constants $B_1$ and $B_2$ with
$$ c^{(n+2)} \leq c^{(n)} \left( \frac{n-2}{n} + \frac{B_1}{n(n-1)} \right) + \frac{B_2}{n},$$
hence
$$ c^{(n+2)} - c^{(n)} \leq \frac{1}{n} \left(\left( -2 + \frac{B_1}{n-1}\right) c^{(n)} + B_2 \right).$$
Now let $n-1 > B_1$. Then for $c^{(n)} > B_2$, the above
inequality shows $c^{(n+2)} < c^{(n)}$, while for $c^{(n)} \leq
B_2$, $c^{(n+2)} \leq c^{(n)} + B_2/n \leq B_2 (1 + 1/n).$ Thus
$c^{(n)}$ is a bounded sequence. \qedi
\end{proof}

\begin{remark}
We will make no use of the fact that the logarithmic correction to
the $1/n$-decay of the higher coefficients occurs only in the
coefficient $a_1^{(n)}$. We chose to include this in the statement
of the preceding theorem anyway, because this gives some insight
into the nature of the recursion and is not hard to prove.
\end{remark}
\begin{remark} Numerical calculations of the first few thousand
$a_j^{(n)}$ suggest that (c) above continues to be true if we
choose $p=1$, but this seems to be much harder to prove. However,
the estimate above is more than good enough for us.
\end{remark}
\begin{remark}
The constants appearing in the proof of Proposition~\ref{a_n
limit} (b) and (c) are not optimal, and could be improved by more
careful arguments. This is unimportant for our purposes, and for
the sake of brevity and readability we chose to use the simple
estimates given.
\end{remark}

\begin{corollary} \label{b_n limit}
Let $b_j^{(n)}$ be given by (\ref{bj in proof}). Then for each $p
> 1$, there exists $C_3>0$ such that
$$ \sup_{j \geq 0} p^{-j} b_j^{(n)} \leq \frac{C_3}{n-1}.$$
\end{corollary}

\begin{proof}
For $j \leq n-1$, we have $n-1 \leq j(n-j)$, and thus
Proposition~\ref{a_n limit} (c) gives
\begin{eqnarray*}
p^{-j} b_j^{(n)} & \leq & \frac{p^{-j}}{n-j} \left( (a_0^{(n)} + a_1^{(n)})
+ \frac{C_2}{n-1} \frac{p^2}{p-1} (p^{j-1}-1) \right) \leq \\
& \leq & p^{-j}  \frac{j(a_0^{(n)} + a_1^{(n)})}{n-1} + \frac{p\,
C_2}{p-1} \frac{1}{n-1} \leq \frac{C_3}{n-1}\,.\qquad\qedi
\end{eqnarray*}
\end{proof}
 Having good control over the coefficients
$a_j^{(n)}$, we can now derive relatively sharp  estimates on the
functions $x_n, y_n$ and $z_n$. Let us fix $\alpha < 1$ and define
$$R_n^\alpha(t) = \frac{1}{(n-1)^\alpha} \max \left\{ \left|
\frac{t_{\rm c}}{t+\I t_{\rm c}} \right|^n, \left(
\frac{1}{\sqrt{2}} \right)^{n-2} \left| \frac{t_{\rm c}}{t+\I
t_{\rm c}} \right|^2 \right\}.$$ Obviously, for $t \leq t_{\rm c}$
the first function in the maximum above dominates, for $t>t_{\rm
c}$ the second one does.

For families of functions $g_n(t), G_n(t)$ we write
$$ g_n(t) = \Or(G_n(t))$$
if there exists $C>0$ such that $|g_n(t)| \leq C |G_n(t)|$ for all
$n \in \N$ and all $t \in \R$.
\begin{theorem} \label{main estimates}
For $n>1$ and $\alpha < 1$, we have
\begin{eqnarray}
x_n(t) & = & \I \frac{(n-1)!}{t_{\rm c}^n} \left( \frac{2\sin(\gamma\pi/2)}{\pi} \tRe
\left(\left( 1 - \I \frac{t}{t_{\rm c}} \right)^{-n}\right) + \Or(R_n^\alpha(t)) \right), \label{AA}\\
y_n(t) & = & \frac{(n-1)!}{t_{\rm c}^n} \,\,\Or(R_n^\alpha(t)), \label{BB}\\
z_n(t) & = & - \frac{(n-1)!}{t_{\rm c}^n} \left(
\frac{2\sin(\gamma\pi/2)}{\pi} \tIm \left(\left( 1 - \I
\frac{t}{t_{\rm c}} \right)^{-n}\right) + \Or(R_n^\alpha(t))
\right), \label{AB}\
\end{eqnarray}
\end{theorem}

\begin{proof}
With the definition of $f$ and Proposition~\ref{a_n limit} (a) we
get
$$ a_0^{(n)} \tIm(f^n) = \frac{2 \sin(\gamma\pi/2)}{\pi\gamma} \tIm
\left(\left( 1 - \I \frac{t}{t_{\rm c}} \right)^{-n}\right) + O\left( \frac{1}{n^2}
 \left|\frac{t_{\rm c}}{t+\I t_{\rm c}} \right|^n\right)$$
when $n$ is even, and a similar formula for $a_0^{(n+1)}
\tRe(f^n)$ when $n$ is odd. This covers the $j=0$ terms in
(\ref{x_n recursion step}) and (\ref{z_n recursion step}). For the
remaining terms, let
$$c_j^{(n)}  =  \left\{         \begin{array}{ll}
                    a_j^{(n)} & \,\,\mbox{if } n \mbox{ is even,} \\
                    n a_j^{(n)}/(n-j) &\,\, \mbox{if } n \mbox{ is odd.}
                    \end{array} \right.$$
Now $n/(n-j) \leq j$ for $j < n$, and thus by Proposition~\ref{a_n
limit} (b) and  (c) for each $p>1$ we can find $C>0$ such that
$$c_j^{(n)} \leq j p^j \frac{C}{(n-1)^{\alpha}} $$
for all $j \geq 1$. (For $j \geq 2$, we may even choose $\alpha =
1$, but we will not exploit this.) For $|t| \leq t_{\rm c}$, we
have $|t_{\rm c}/(t+\I t_{\rm c})|^{-j} \leq 2^{j/2}$, so we get
$$ \left| \sum_{j=2}^{n-1} \frac{c_j^{(n)}}{2^j} \left( \frac{\I t_{\rm c}}{t+\I t_{\rm c}}\right)^{n-j}
\right| \leq \left( \frac{C}{(n-1)^\alpha}  \sum_{j=2}^{n-1} j \left(\frac{p}{\sqrt{2}} \right)^j \right)
 \left| \frac{t_{\rm c}}{t+\I t_{\rm c}} \right|^n.$$
If we choose $p < \sqrt{2}$, the sum on the right hand sided is
bounded uniformly in $n$.  Combining this with our above
calculations, (\ref{AA}) and (\ref{AB}) are proved for $|t| <
t_{\rm c}$. For $|t| > t_{\rm c}$, we have $|t_{\rm c} / (t+\I
t_{\rm c})| \leq 1/\sqrt{2}$, and thus
$$
 \left| \sum_{j=2}^{n-2} \frac{c_j^{(n)}}{2^j} \left( \frac{\I t_{\rm c}}{t+\I t_{\rm c}}\right)^{n-j} \right|
 \leq  \frac{C}{(n-1)^\alpha} \left| \frac{t_{\rm c}}{t+\I t_{\rm c}} \right|^2  \sum_{j=2}^{n-2} j
            \left(\frac{p}{2}\right)^j \left( \frac{1}{\sqrt{2}} \right)^{n-2-j}. $$
If we choose again $p < \sqrt{2}$, the sum on the right hand side
is bounded by $\tilde{C}(1/\sqrt{2})^{n-2}$ uniformly in $n$. For
the term with $j = n-1$, this does not work since then $n-2-j <
0$. But for $n$ even, this term vanishes since then $c_{n-1}^{(n)}
= 0$, and for $n$ odd, it equals
$$ \frac{c_{n-1}^{(n)}}{2^{n}} \tRe\left( \frac{\I t_{\rm c}}{t+\I t_{\rm c}} \right) =
 \frac{n a_{n-1}^{(n+1)}}{2^n} \frac{t_{\rm c}^2}{t^2+t_{\rm c}^2} \leq \frac{\tilde C}{n-1}
  \left(\frac{1}{\sqrt{2}}\right)^{n-2} \left| \frac{t_{\rm c}}{t+\I t_{\rm c}} \right|^2.$$
This proves (\ref{AA}) and (\ref{AB}) for $|t| \geq t_{\rm c}$.
The proof of (\ref{BB}) is similar and uses Corollary \ref{b_n
limit}. \qedi
\end{proof}


\section{Optimal truncation}
By the results of the previous section $\pi_k$ grows like
$(k-1)!/t_{\rm c}^k$. Hence, the sum $\pi^{(n)} = \sum_{k=0}^n
\eps^k \pi_k$ does not converge to an exactly equivariant
projection $\pi^{(\infty)}$ as $n \to \infty$. This is the reason
why we see exponentially small transitions. The basis in which
these transitions develop smoothly is the optimal superadiabatic
basis: since we cannot go all the way to infinity with $n$, we fix
$\eps$ and choose $n=n(\eps)$ such that the off-diagonal elements
in (\ref{general diag}) become minimal. Using Stirling's formula
and (\ref{AA}) resp.\ (\ref{AB}), it is easy to see that the place
to truncate is at $n(\eps) = t_{\rm c}/ \eps$. This $n(\eps)$ is
in general not a natural number, but we will find that a change of
$n$ which is of order one does not change the results. Before we
go into more details, we need a preliminary result.

\begin{lemma} \label{power to exponential}
Uniformly in $x \in [0,1]$ and for $k > 0$, we have
$$ (1+x)^{-k} = \e^{-k x} + \e^{-kx/2} \Or{\textstyle \left(\frac{1}{k}\right)}.$$
\end{lemma}
\begin{proof}
We start with the equality
\begin{equation} \label{diff1}
 (1+x)^{-k} - \e^{-kx} = \e^{-kx} \left(\e^{k(x-\ln(1+x))} -1 \right).
 \end{equation}
At first consider $x > \sqrt{1/k}$. There we use the inequality
 $(x-\ln(1+x)) \leq  x/3$, valid for $0 \leq x \leq 1$, in (\ref{diff1}) and obtain
$$
| (1+x)^{-k} - \e^{-kx} | \leq \e^{-kx} \left( \e^{kx/3} - 1
\right) = \e^{-kx/2} \left( \e^{-kx/6} -   \e^{-kx/2}\right).
$$
For $x> \sqrt{1/k}$, the term in the last bracket above is
$\Or(1/k)$, and we are done in this case. For $x\leq\sqrt{1/k}$,
we use $(x-\ln(1+x)) \leq x^2/2$ and rearrange (\ref{diff1}) to
get
$$
\e^{kx/2} ((1+x)^{-k} - \e^{-kx}) = \e^{-kx/2} \left(\e^{kx^2/2}
-1 \right) =: f(x,k).
$$
To find out where $f(x,k)$ is maximal, we calculate
$$ \frac{\D}{\D x}f(x,k) = \frac{k}{2} \e^{-kx/2} \left( 1 + \e^{kx^2/2}(2x-1)\right).$$
The derivative is zero exactly at the solutions of the equation
\begin{equation} \label{nullst}
\ln(1-2x)/x^2 = -k/2.
\end{equation}
Now $\ln(1-2x)/x^2 = -2/x + R(x)$, where $R(x)$ is a power series
in $x$, convergent for $x < 1/2$. Thus for $x < \sqrt{1/k}$ and
$k$ sufficiently large, there exists exactly one solution
$x^{\ast}(k)$ of (\ref{nullst}), and $x^{\ast}(k) < C/k$ uniformly
in $k$ for some $C>0$. Since $\frac{\D}{\D x}f(x,k) > 0$ for $x <
1/k^2$, $f(x,k)$ has a maximum at $x^{\ast}(k)$. Thus
$$ f(x,k) \leq f(x^{\ast}(k),k) \leq \e^{-C/2k}-1 = \Or(1/k)$$
for $x < \sqrt{1/k}$, and the claim is proved. \qedi
\end{proof}

Lemma \ref{power to exponential} immediately yields
\begin{equation} \label{R1neu}
\left(1+\frac{a}{k}\right)^{-k} = \e^{-a}
\,\left(1+\Or\left({\textstyle \frac{1}{k}}\right)\right)
\end{equation}
uniformly on compact intervals of $a$ by taking $x = a/k$.

We now turn to the proof of Theorem \ref{MainThm}, which we deduce
from Theorems \ref{general diag} and \ref{main estimates}. As
stated already in (\ref{ndef}), we will use
\begin{equation}\label{ndef2}
n_\eps = \frac{t_{\rm c}}{\eps} -1+ \sigma_\eps,
\end{equation}
where $\sigma_\eps \in [0,2[$ is such that $n_{\eps}$ is an even
integer. The advantage of this convention about $\sigma_\eps$ is
that now the off-diagonal components in (\ref{general offdiag eq})
are always given by $\eps^{n_\eps+1}x_{n_\eps+1}$ since
$z_{n+1}=0$ for even $n$. Of course we could as well consider the
asymptotic behavior of $\eps^{n+1}z_{n+1}$ for odd $n$ and one
would expect to end up with the same result. However, it is
obvious from (\ref{AA}) and (\ref{AB}) that $x_{n+1}$ is purely
imaginary and $z_{n+1}$ is real at leading order. Thus the large
$n$ asymptotics of the off-diagonal component of the effective
Hamiltonian do depend on whether we consider even or odd
superadiabatic bases. On the other hand, the asymptotics of the
propagator must be independent of the exact choice of basis. We
will discuss this point after giving the proof of
Theorem~\ref{MainThm} based on the above convention.

\begin{proof}[Proof of Theorem~\ref{MainThm}]
 We want to apply Theorem \ref{general
diag} and thus have to check that $\xi, \eta$ and $\zeta$ defined
in (\ref{xi})--(\ref{zeta}) together with their derivatives are
$\Or(\eps \theta')$. From Proposition~\ref{z_n and recursion}
together with Proposition~\ref{a_n limit} we infer that there
exists $C>0$ such that $|x_k(t)| \leq C \theta'(t) (k-1)!/ t_{\rm
c}^{k}$ for each $k$. The same is true for $y_n$ and $z_n$. Using
the differential equations (\ref{E1a})--(\ref{E1c}), we find that
there is $C' > 0$ with $|x_n'(t)| \leq C' \theta'(t) n!/ t_{\rm
c}^{n+1}.$ This means that
$$ |\xi'(t)| \leq \eps C' \theta'(t) \sum_{k=1}^n \eps^k t_{\rm c}^{-k-1} k! \,\eps^{k-1},$$
with similar expressions for the other quantities. Now taking
$\eps = t_{\rm c}/(n_\eps - \sigma_\eps)$, we find
\[
\sum_{k=0}^{n_\eps} t_{\rm c}^{-k} \eps^k (k+1)! =
\sum_{k=0}^{n_\eps} \frac{(k+1)!}{(n_\eps-\sigma_\eps)^k}
 =   \left( 1 + \frac{2}{n_\eps-\sigma_\eps} +
\frac{3!}{(n_\eps-\sigma_\eps)^2} + \ldots \right).
\]
Each of the $n_\eps+1$ terms in the sum above is bounded by
$\mathrm{const}/(n_\eps-\sigma_\eps)$ except the first which is
$1$. This shows
$$ |\xi'(t)| \leq \eps \theta' C' {\textstyle\left(1 + \frac{n_\eps}{n_\eps-\sigma_\eps}\right)}\,,$$
and Theorem \ref{general diag} gives (\ref{Hop}) with
$c_\eps^{\op}(t) = \eps^{n_\eps+1}x_{n_\eps+1}(t)(1+\Or(\eps
\theta'(t))$. Recall  that $z_{n_\eps+1}(t) = 0$ due to our
convention. It remains to determine the leading order asymptotics
of $\eps^{n_\eps+1}x_{n_\eps+1}$. For convenience of the reader
let us rewrite (\ref{AA}) as
\begin{equation}\label{epsx}
\eps^{n_\eps+1}x_{n_\eps+1}(t)  = \I
\frac{\eps^{n_\eps+1}n_\eps!}{t_{\rm c}^{n_\eps+1}}
{\textstyle\left[ \frac{2\sin(\gamma\pi/2)}{\pi} \,\tRe
\left(\hspace{-2pt}\left( 1 - \I \frac{t}{t_{\rm c}}
\right)^{-(n_\eps+1)}\right) +
\Or\left(R_{n_\eps+1}^\beta(t)\right)\hspace{-1pt} \right]}.
\end{equation}

\begin{lemma} \label{stirling}
With (\ref{ndef2}), we have
$$ \frac{\eps^{n_\eps+1}n_\eps!}{ t_{\rm c}^{n_\eps+1}}= \sqrt{\frac{2 \pi \eps}{t_{\rm c}}}
 \e^{-\frac{t_{\rm c}}{\eps}}(1 + \Or(\eps)).$$
\end{lemma}
\begin{proof}
Stirling's formula for $(n+1)!$ implies
$$ n! =  \frac{1}{n+1}\left( \frac{n+1}{\e} \right)^{n+1} \sqrt{n+1}
\sqrt{2 \pi}\,\left(1+\Or\left({\textstyle \frac{1}{n+1}}\right)\right)\,.$$
Together with (\ref{R1neu}) this yields
\begin{eqnarray*}
\eps^{n_\eps+1} n_\eps! & = & t_{\rm c}^{n_\eps+1}\,
\e^{-(n_\eps+1)} \left( 1 - \frac{\sigma_\eps}{n_\eps+1}
\right)^{-(n_\eps+1)}
\sqrt{\frac{2\pi}{n_\eps+1}}\,\left(1+\Or\left({\textstyle \frac{1}{n_\eps+1}}\right)\right) = \\
& = & t_{\rm c}^{n_\eps+1}\,
\e^{-(n_\eps+1)}\,\e^{\sigma_\eps} \,\sqrt{\frac{2\pi}{n_\eps+1}}\,
\left(1+\Or\left({\textstyle \frac{1}{n_\eps+1}}\right)\right)  = \\
& = & t_{\rm c}^{n_\eps+1}\, \e^{-\frac{t_{\rm c}}{\eps}}\,
\sqrt{\frac{2 \pi \eps}{t_{\rm c} + \eps\sigma_\eps }}\,(1 +
\Or(\eps)).
\end{eqnarray*}
Finally,
$$\sqrt{\frac{2 \pi \eps}{t_{\rm c} + \eps\sigma_\eps }} = \sqrt{\frac{2 \pi \eps}{t_{\rm c}}}
\left(1+ \frac{\eps\sigma_\eps}{t_{\rm c}} \right)^{-1/2} =
\sqrt{\frac{2 \pi \eps}{t_{\rm c}}}\,(1+\Or(\eps))\,.\qquad\qedi
$$
\end{proof}

Lemma~\ref{stirling} takes care of the first factor in
(\ref{epsx}). Turning to the terms inside the square brackets in
(\ref{epsx}), let us first note that for $|t| \geq t_c$, both
terms are $$\Or(2^{-(n_\eps-1)/2}/(1+t^2) )= \Or(\exp(-t_{\rm c}
\ln 2/(2 \eps))/(1+t^2))\,,$$ proving the theorem in this case.
For $|t| < t_c$, we investigate the modulus and the phase
separately. Let $0<\beta<1$. From Lemma \ref{power to exponential}
it follows that
\[
 {\textstyle\left| 1 + \I \frac{t}{t_{\rm c}} \right|}^{n_{\eps}+1} =
 {\textstyle\left(1+ \frac{t^2}{t_{\mr c}^2}\right)}^{(t_{\mr c}/ \eps + \sigma_\eps)/2}
 ={\textstyle \left(1+ \frac{t^2}{t_{\mr c}^2}\right)}^{\sigma_\eps/2} \left( \e^{-\frac{t^2}{2 t_{\mr c} \eps}} +
 \Or \left( \eps \e^{-\frac{t^2}{4 t_{\mr c} \eps}} \right) \right).
\]
For $|t| \geq \eps^{\beta/2}$, $\exp(-t^2/(2 t_{\mr c} \eps)) =
\Or(\eps \exp(-t^2/(4 t_{\mr c} \eps)))$. Thus neither the
prefactor involving $\sigma_\eps$ above nor the phase play any
role in this region. For $|t| < \eps^{\beta/2}$, $(1+ t^2/t_{\mr
c}^2)^{\sigma_\eps/2} = 1 + \Or(\sigma_\eps \eps^\beta)$ and
therefore
$$ {\textstyle\left| 1 + \I \frac{t}{t_{\rm c}} \right|}^{n_{\eps}+1} = \e^{-\frac{t^2}{2 t_{\mr c} \eps}} +
 \Or \left( \eps^\beta \e^{-\frac{t^2}{4 t_{\mr c} \eps}} \right).$$
 The same reasoning applies to $R_{n_\eps+1}^\beta$ and gives
 $$ R_{n_\eps+1}^\beta(t) \leq \eps^\beta \left( \e^{-\frac{t^2}{2 t_{\rm c} \eps}} +
 \Or\left(\eps^\beta \e^{-\frac{t^2}{4 t_{\rm c} \eps}} \right)\right). $$
Turning to the phase in the region $|t| < \eps^{\beta/2}$, we find
\begin{eqnarray*}
\e^{\I (n_\eps+1)\arctan (t/t_{\rm c})} &=& \exp \left( \I
\left({\textstyle \frac{t_{\rm c}}{\eps}} + \sigma_\eps \right)
\left(
(t/t_{\rm c}) -{\textstyle \frac{1}{3}}(t/t_{\rm c})^3 + \Or((t/t_{\rm c})^5)\right) \right) = \\
& = & \exp \left( \I \left( {\textstyle \frac{t}{\eps}} -
{\textstyle \frac{t^3}{3\eps t_{\rm c}^2} } +
{\textstyle\frac{\sigma_\eps t}{t_{\rm c}}}\right)+ \Or(t_{\rm c}
(t/t_{\rm c})^5/\eps) + \Or(\sigma_\eps (t/t_{\rm c})^3)
\right)\\
& =& \exp \left( \I \left( {\textstyle \frac{t}{\eps}} -
{\textstyle\frac{t^3}{3\eps t_{\rm c}^2}} +
{\textstyle\frac{\sigma_\eps t}{t_{\rm c}}}\right)\right) \left(1
+ \Or(\eps^{5\beta /2 - 1}) + \Or(\eps^{3\beta/2})\right).
\end{eqnarray*}
Now we just have to collect all the pieces and add the complex
conjugate. \qedi
\end{proof}

Let us now see what of the above would have changed for $n_{\eps}$
odd. Then $x_{n_\eps+1} = 0$, and (\ref{AB}) together with Lemma
\ref{power to exponential} and \ref{stirling} yields
\begin{eqnarray} \label{zeta eps}
c_\eps^{n_\eps}(t) &=& -\eps^{n_\eps+1}z_{n_\eps+1}(t)\, (1 +
\Or(\eps\theta'))\\& =& 2\,\sqrt{\frac{2\eps}{\pi t_{\rm
c}}}\,\sin\left(\frac{\pi \gamma}{2}\right)\, \e^{-\frac{t_{\rm
c}}{\eps}}\,\e^{-\frac{t^2}{2\eps t_{\rm c}}}
\,\sin\left(\frac{t}{\eps}-\frac{t^3}{3\eps t_{\rm c}^2} +
\frac{\sigma_\eps t}{t_{\rm c}} \right)+
\Or\left(\phi^\alpha(\eps,t)\right)\,.\nonumber
\end{eqnarray}
At first, this looks like an important difference, since now the
off-diagonal elements in the transformed Hamiltonian are purely
real-valued in leading order, while in the other case they were
purely imaginary. However, in the computation of the propagator,
another factor of $\exp(\pm \I t/\eps)$ from the dynamical phase
appears, cf.\ (\ref{oscint}). At leading order only the resonant
term of the Hamiltonian survives, which is the same for odd and
even $n_\eps$.

\section{First order perturbation in the optimal
superadiabatic basis}

In this section we prove Corollary~\ref{PropCor}. Since we use
standard first order perturbation theory, we stay sketchy in some
parts. After splitting $H_\eps^{\op}(t)$, see (\ref{Hop}), as
\[
H_\eps^{\op}(t) =\left(
\begin{array}{cc}\frac{1}{2}&0\\0&-\frac{1}{2}\end{array}\right) +
V_\eps(t) =: H_0 + V_\eps(t)\,,
\]
Dyson expansion in the interaction picture (cf.\ \cite{ReSi2},
Thm.\ X.69) yields
\begin{eqnarray*}
K_\eps^{\op}(t,s) &=& \e^{-\frac{\I t H_0}{\eps}} \left( {\rm id}
-\frac{\I}{\eps}\int_s^t\e^{\frac{\I \tau
H_0}{\eps}}\,V_\eps(\tau)\,\e^{-\frac{\I \tau H_0}{\eps}}\,\D
\tau\right)\e^{\frac{\I s H_0}{\eps}}\\ &&+ \,\left(
\begin{array}{cc} \Or(\eps^2) & \Or(\eps\e^{-\frac{t_{\rm
c}}{\eps}})
\\\Or(\eps\e^{-\frac{t_{\rm c}}{\eps}})&\Or(\eps^2)
\end{array}\right) \Delta(t,s)\,.
\end{eqnarray*}
Thus we only need to evaluate the integral
\begin{eqnarray*}\nonumber
-\frac{\I}{\eps}\int_s^t\e^{\frac{\I \tau
H_0}{\eps}}\,V_\eps(\tau)\,\e^{-\frac{\I \tau H_0}{\eps}}\,\D \tau
&=&-\frac{\I}{\eps}\int_s^t
\left(\begin{array}{cc}0&\e^{\frac{\I\tau}{\eps}}
c_\eps^{\op}(\tau)\\ \e^{-\frac{\I\tau}{\eps}}
\overline{c}_\eps^{\op}(\tau) &0 \end{array}\right)\,\D \tau\\&&\,
+\left(\begin{array}{cc}\Or(\eps)&0\\0&\Or(\eps)\end{array}\right)\Delta(t,s)\,.
\end{eqnarray*}
Inserting (\ref{Hod}) and using (\ref{Phialpha}) gives
\begin{eqnarray}\nonumber\lefteqn{
-\frac{\I}{\eps}\int_s^t\e^{\frac{\I\tau}{\eps}}
c_\eps^{\op}(\tau)\D\tau=}\\ \nonumber&=& \sqrt{\frac{2}{\eps\pi
t_{\rm c}}}\,\sin\left(\frac{\pi \gamma}{2}\right)\,
\e^{-\frac{t_{\rm c}}{\eps}} \int_s^t\e^{\frac{\I\tau}{\eps}}
\,\e^{-\frac{\tau^2}{2\eps t_{\rm c}}} \,\left(\e^{-\frac{\I
\tau}{\eps} +\frac{\I\tau^3}{3\eps t_{\rm c}^2} - \frac{\I \sigma
\tau}{t_{\rm c}} }+ \e^{\frac{\I \tau}{\eps}
-\frac{\I\tau^3}{3\eps t_{\rm c}^2} + \frac{\I \sigma \tau}{t_{\rm
c}}  } \right) \D\tau
\\ & & +\,
\Or\left(\eps^\alpha\e^{-\frac{t_{\rm c}}{\eps}}\Delta(t,s)\right)
= (*)\,,\label{oscint}
\end{eqnarray}
for each $\alpha<1$.  Now we replace the exponentials
$\e^{\pm(\frac{\I\tau^3}{3\eps t_{\rm c}^2} - \frac{\I \sigma
\tau}{t_{\rm c}} )}$ by $1\pm(\frac{\I\tau^3}{3\eps t_{\rm c}^2} -
\frac{\I \sigma \tau}{t_{\rm c}} ) $. Using
$|\e^{\I\varphi}-1-\I\varphi|\leq \varphi^2$, we conclude that the
resulting error is bounded by a constant times
\[
 \eps^{-\frac{1}{2}}
\e^{-\frac{t_{\rm c}}{\eps}} \int_{-\infty}^\infty
\,\e^{-\frac{\tau^2}{2\eps t_{\rm c}}}
\left(\frac{\tau^6}{\eps^2}+\frac{\tau^4}{\eps} +\tau^2
\right)\D\tau= \Or(\eps\e^{-\frac{t_{\rm c}}{\eps}} )\,.
\]
 Hence we obtain
\begin{eqnarray}\nonumber
(*) &=& {\textstyle\sqrt{\frac{2}{\eps\pi t_{\rm
c}}}\,\sin\left(\frac{\pi \gamma}{2}\right)\, \e^{-\frac{t_{\rm
c}}{\eps}}}\hspace{-1pt} \int_s^t
\hspace{-1pt}\e^{-\frac{\tau^2}{2\eps t_{\rm c}}}
\,\left(1+{\textstyle\frac{\I\tau^3}{3\eps t_{\rm c}^2} - \frac{\I
\sigma \tau}{t_{\rm c}}}+ \e^{\frac{2\I \tau}{\eps}
 } {\textstyle \left(1-\frac{\I\tau^3}{3\eps t_{\rm c}^2} + \frac{\I \sigma
\tau}{t_{\rm c}}\right)}\right)\hspace{-1pt} \D\tau\\&& +\,
\Or\left(\eps^\alpha\e^{-\frac{t_{\rm
c}}{\eps}}\Delta(t,s)\right)\label{oscint2}
\end{eqnarray}
with $\alpha<1$, where the first summand in the integrand gives
rise to the explicit term in (\ref{Koffdiag}). The remaining terms
can be integrated explicitly as well, most conveniently using
Maple or Mathematica. They are all of order
$\Or(\sqrt{\eps}\e^{-\frac{t_{\rm c}}{\eps}}\Delta(t,s))$
uniformly in $t$ and $s$ resp.\ of order
$\Or(\eps^\alpha\e^{-\frac{t_{\rm c}}{\eps}}\Delta(t,s))$ for
$|t|$ and $|s|$ larger than $\eps^\beta$ for some
$\beta<\frac{1}{2}$. To illustrate the reasoning note that
\[
\int_s^t \e^{-\frac{\tau^2}{2\eps t_{\rm c}}}\tau\,\D\tau = \eps
t_{\rm c}\left(\e^{-\frac{s^2}{2\eps t_{\rm
c}}}-\e^{-\frac{t^2}{2\eps t_{\rm c}}}\right)\,.
\]
This is uniformly of order $\Or(\eps)$, but of order
$\Or(\e^{-\eps^{2\beta-1}})$ for $|t|$ and $|s|$ larger than
$\eps^\beta$. Finally we emphasize  that we could get the next
order corrections to (\ref{Koffdiag}) explicitly by evaluating
(\ref{oscint2}).

\end{document}